\documentclass{revtex4}

\usepackage{graphicx}
\usepackage{aas_macros}
\input epsf

\def\apjl{Astrophys. J. Lett.}
\def\be{\begin{equation}}
\def\ee{\end{equation}}
\def\bea{\begin{eqnarray}}
\def\eea{\end{eqnarray}}

\def\cmm2{{\,\rm cm^{-2}}}
\def\cm2{{\,{\rm cm}^2}}
\def\cmm3{{\,{\rm cm}^{-3}}}
\def\gcmm3{{\,{\rm g\,cm^{-3}}}}

\def\fun#1#2{\lower3.6pt\vbox{\baselineskip0pt\lineskip.9pt
  \ialign{$\mathsurround=0pt#1\hfil##\hfil$\crcr#2\crcr\sim\crcr}}}

\def\p3m{P$^3$M}

\def\la{\mathrel{\mathpalette\fun <}}

\def\fun#1#2{\lower3.6pt\vbox{\baselineskip0pt\lineskip.9pt
  \ialign{$\mathsurround=0pt#1\hfil##\hfil$\crcr#2\crcr\sim\crcr}}}
\begin{document}
% You should use BibTeX and revtex.bst for references
\bibliographystyle{prsty}

% number  and your conference paper identification number on the
% title page in preprint mode. Multiple \preprint commands are allowed.
%\preprint{}

\title{CMB and Inflation:  the report from Snowmass 2001}
% Optional argument for running titles on pages
%\title[]{}



\author{Sarah Church}
\affiliation{Department of Physics, Stanford University, Stanford CA
94305-4060} \email{schurch@stanford.edu}
\author{Andrew Jaffe}
\affiliation{UC Berkeley; Imperial College, London}
\email{a.jaffe@ic.ac.uk}
\author{Lloyd Knox}
\affiliation{UC Davis}
\email{lknox@ucdavis.edu}
%\thanks{}


\date{\today}

\begin{abstract}
Astrophysical observations provide the best evidence of physics beyond the
particle physicists' standard model.  Over the past decade, the case has
solidified for inflation, dark energy, dark matter and massive neutrinos.
In this report we focus on inflation and dark energy as constrained by the
CMB---first the evidence from current observations and then what we can
expect to learn in the future. High energy physicists can contribute to
the succesful realization of this future, both observationally and
theoretically.
\end{abstract}
% \pacs{}

%\maketitle must follow title, authors, abstract and \pacs
\maketitle

%\section{}
%\label{}
%\subsection{}
%\subsubsection{}

\section{Introduction}
\label{introduction}

Accelerator--based efforts spanning decades have revealed no
direct evidence of physics beyond the Standard Model. 
Although this is likely to
change with the Large Hadron Collider (LHC) or possibly even with
the current (higher--luminosity) run at the Tevatron, we point out
that {\em astrophysical} evidence for physics beyond the standard
model has already grown to be conclusive. Observations indicate
the need for
\begin{itemize}
\item inflation,
\item dark energy,
\item dark matter, and
\item massive neutrinos.
\end{itemize}
In short, cosmology is delivering on its promise of probing fundamental
physics.

These results are creating intellectual excitement in the high
energy physics community. They also signal an opportunity.  The
prospects are bright for further progress in our understanding of
fundamental physics, via cosmology, and researchers trained in
particle physics should not only be interested, but are
well--equipped to contribute. Indeed, a number of particle
physicists have become actively engaged in astrophysical
observations.

In this report we review the current status of observations of the
temperature of the cosmic microwave background (CMB), and explain
how they provide evidence of an inflationary era of expansion in
the early Universe and the existence of ``dark energy''.  One of
our intentions in providing this review is to demonstrate that
cosmology has a track record which suggests that we have some
understanding of the system we are studying.  This track record
bolsters the case that we will be able to get meaningful answers
to the more detailed questions that are motivating planned
observations.  These future observations, which we discuss,
include more detailed measurement of temperature maps, detection
and measurement of polarization (including the modes generated by
gravitational waves) and Sunyaev--Zeldovich surveys.  Of course,
observational cosmology is much broader than the topics we are
covering, and the interested reader is encouraged to consult the
other P4 and E6 subtopic reports.

\section{Overview of CMB and Fundamental Physics}
\label{overview}

Observations of the CMB are powerful probes of physical
processes in the early Universe.  This is due to the
richness of the observables and the relative ease
with which they can be calculated from first principles.
The CMB may be unique in astrophysics in that its properties,
over a wide range of angular scales, can be accurately
predicted for a given model in linear perturbation theory.

The CMB is most importantly a probe of structure
formation in the Universe.  The past decade has confirmed
that structure in the Universe grew from highly--uniform
initial conditions (density fluctuations of order 1 part
in $10^5$) via gravitational instability.
The evolution of these initial perturbations is sensitive
to a number of cosmological parameters, including the
mean curvature, density of baryons, density of cold dark matter,
density of hot dark matter, density of dark energy, and the
epoch of reionization of the IGM.  Observations of CMB anisotropy
are thus sensitive to these parameters and, in addition, the statistical
properties of the initial conditions.

We do not expect these ``initial conditions'' to be genuinely
initial but rather the result of some dynamical process. The CMB
is already providing us strong evidence that this generation
occurred during an inflationary stage in the expansion of the
Universe. The fundamental physics responsible for inflation is far
from being understood, although there is no dearth of toy models.
The possible energy scale of this phenomenon ranges from $10^3$
GeV up to $10^{16}$ GeV.

The applicability of linear perturbation theory is due to
the fact that the microwave background photons last interacted
with matter at a redshift of $z \simeq 1100$ when the perturbations
were still well in the linear regime.  We can probe structure
at this epoch by making maps of the intensity of the CMB
(or, equivalently its temperature) and also of the polarization.
We have already learned much from temperature maps (to be summarized
below) and expect the first detections of polarization within the
next one to two years.
We expect high sensitivity, high angular resolution maps of temperature
and polarization to provide strong constraints on models of inflation,
and high--precision constraints on cosmological parameters.

High--sensitivity polarization maps may actually allow us to
determine the energy scale of inflation.  Although both scalar
(density) perturbations and tensor (gravitational wave)
perturbations to the metric tensor result in curl--free
polarization patterns, only the tensor perturbations result in
non--zero curl.  Detection of curl in the polarization pattern
(sometimes referred to as the ``B mode'') would be evidence for
gravitational waves and their amplitude is directly proportional
to the energy--scale of inflation.  Detecting the B--mode is a
very challenging task, probably requiring large--scale dedicated
detector arrays (either ground or space--based).  We consider succesful 
detection of this signal to be very exciting but a long shot.  Further
observations will shed light on our chances of success.

Observations of the last--scattering surface do not provide strong
constraints on the nature of the dark energy (such as its density today,
and equation of state).  Fortunately, not all CMB photons arrive at us
today undisturbed by matter along the line--of--sight.  Some pass
through rich clusters of galaxies whose hot electrons scatter them to
higher energies, creating a spectral distortion called the
Sunyaev--Zeldovich (SZ) effect.  Since the number density of clusters is
sensitive to the dark energy density and equation of state, SZ surveys
can tell us about these quantities.  Gravitational lensing of the CMB and
the ``Rees--Sciama'' effect also create anisotropies at late times with
statistical properties that are sensitive to the nature of the dark energy.

To summarize this overview, there are three observable properties of the
CMB which all provide unique constraints on particle physics models. These
are the spectrum (observed to be that of a black body, except in the
direction of clusters of galaxies), temperature anisotropy, and
polarization.  In Sections~\ref{inflation} and~\ref{acoustic} we review
the physics probed by the CMB, and in Section~\ref{current} we report on
the current status of temperature and polarization measurements. In
section~\ref{motivation} we discuss the future of CMB temperature and
polarization measurements, in section~\ref{analysis} we discuss the
challenges of analyzing the large data sets expected from future
experiments, and in section~\ref{sz_v4} we review the ways in which SZ
surveys probe cosmological parameters, including dark energy.

\section{Inflation and Perturbation Generation}
\label{inflation}

There is a large literature on inflation.  We give a {\em very} brief
treatment here and point the reader to some recent reviews
\cite{linde00,guth00}.

%% Acceleration requires negative pressure
For spatially flat models, the expansion rate and its rate of change are
given by \be H \equiv {\dot a \over a} = \sqrt{8 \pi G \rho \over 3} \ \ ;
\ \ {\ddot a \over a} = -{4 \pi G \over 3}\left(\rho + 3P\right). \ee Thus
acceleration of the scale factor, the essence of inflation, requires $P <
-\rho/3$.

%% Scalar field can do this
Inflation is often modeled as being driven by the vacuum energy of a
spatially homogeneous scalar field, $\phi$.  The dynamics of this scalar field
are usually calculated classically, by use of an effective potential to
account for the quantum--mechanical radiative corrections, $V(\phi)$.  The
equation of motion for this scalar field in an expanding Universe is then
\be
\ddot \phi + 3 H \dot \phi + V'(\phi) = 0
\ee
and the energy density and pressure are:
\bea
\rho & = & V(\phi) + \dot \phi^2/2 \nonumber \\
P &=& - V(\phi) + \dot \phi^2/2
\eea
Inflation occurs when $\dot \phi^2/2 << V(\phi)$ so that $P \simeq - \rho$.

Perturbations are generated quantum--mechanically.  In de Sitter
space, all massless scalar fields have fluctuations with rms
amplitude $\delta \phi = H/(2\pi)$ on the horizon scale.  The resulting
density perturbations are the initial seeds of structure.  There is
a short and simple treatment of the generation and evolution of these density
perturbations in \cite{linde00}.  For a more in depth treatment see
\cite{mukhanov92}.

%% scalar and tensor perturbations
The perturbations in the scalar field result in {\it scalar} metric
perturbations (i.e., they transform like scalars under spatial
rotations).  Inflation also produces {\it tensor} perturbations to the
metric, which are gravitational waves.  Unlike the scalar
perturbations (which are also sensitive to $\partial V(\phi)/ \partial
\phi$) the tensor perturbation amplitude depends only on the expansion
rate during inflation, and thus almost solely on $V(\phi)$.
Detection of these tensor perturbations is necessary for a determination of
the energy scale of inflation.

Inflation faces some conceptual challenges.  For one, the vacuum
energy of the scalar field plays a central role, even though we do not
understand the cosmological constant problem.  Secondly, we have used
an effective potential description which is derived assuming the
scalar field is in a stationary, equilibrium state to describe
{\em dynamics} of the scalar field.  Although there has been no progress
with the first problem, there has been some with the second.  Quantum
corrections to the classical treatment of the evolution of the mean
value of the scalar field, implicit in use of the effective potential,
are only important for potentials with negative curvature, $\partial^2
V(\phi)/ \partial \phi^2<0$.  Even for these models though,
the evolution can be treated in the same
classical manner, but with the tree-level potential replaced
by an effective one with {\em two} scalar fields.~\cite{cormier99} 

\begin{figure}
   \includegraphics[width=4in]{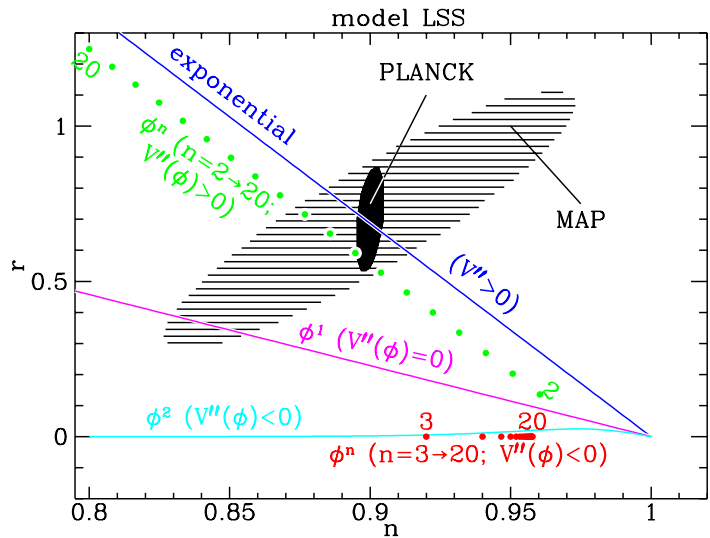}
   \caption[]{Predictions for a range of models in the $n$ and $r$
plane and forecasted error ellipses from expected MAP and Planck data
(from \cite{dodelson97}).}
   \label{lek-infl-fig1}
\end{figure}

%Describe inflation varieties.
Inflation can arise from very simple potentials, including $V(\phi)
\propto \phi^n$ for $n > 1$.  In Figure~\ref{lek-infl-fig1}, reproduced
from \cite{dodelson97}, we show predictions from a variety of
potentials in the plane of power spectrum parameters, $n$ and $r$
where $n$ is the power--law spectral index for the primordial scalar
power spectrum, $P(k) = Ak^n$ and $r \equiv T/S$ where $T$ is the
tensor contribution to the quadrupole anisotropy variance $C_2$ and
$S$ is the scalar contribution.  Also shown are forecasted error
ellipses from expected MAP and Planck data.  Thus it is clear that
precisions measurements of the CMB can lead to great power in
distinguishing among inflationary models.

\section{Physics of Acoustic Oscillations}
\label{acoustic}
After an epoch of inflation, the evolution of the particle species
present in the Universe and of perturbations to densities of those
species is well understood. Here we present a very brief review of this
evolution and the imprint left on the photons that will become the CMB
we observe today. It recaps material presented at greater length and
detail in textbooks (e.g., \cite{KolbTurner,Peebles}) and review
articles (e.g., \cite{HuScoSil94}).

After Big-Bang Nucleosynthesis, the Universe is
composed of relativistic species (photons and neutrinos, with density
$\Omega_r$), baryons in the form of light nuclei ($\Omega_B$), and dark
matter ($\Omega_c$). At temperatures higher than about one eV, the
nuclei remain ionized, and the photons and baryons are tightly coupled
in a plasma. As the Universe cools, the electrons and baryons very
rapidly combine to form (mostly hydrogen) atoms, and the photon
mean-free-path grows to be larger than the scale of the Universe (the
Hubble length) at this time, known as ``Last Scattering,''
``Matter-Radiation Decoupling,'' and ``recombination'' (although
technically these terms refer to slightly different events, they all
occur at roughly the same epoch, at a temperature of $\sim 1$eV or a
redshift of $z\simeq1,100$ or an age of about 300,000 years. [This is
considerably below the naive expectation of 13.6eV due to the
overwhelming number of photons compared to baryons---a factor of
$\sim10^{10}$---keeping the latter ionized longer.]  Free-streaming
through the Universe since this time, these same photons redshift to
become the 2.728K CMB observed today.

The early Universe was highly homogeneous with small density
perturbations, $\delta\equiv\delta\rho/\rho$ in the various components
with density $\rho$. As discussed above, {\em Inflation} is at the
moment the best---and perhaps the only---way of creating these
fluctuations.  The accelerated expansion makes the Hubble length
grow much more slowly than the past horizon, allowing for the causal
creation of fluctuations with wavelengths much larger than the Hubble
length.  

The simplest inflationary models create perturbations where the
fractional perturbations are the same in all species.  Since the
per-particle entropy is therefore spatially constant, such
perturbations are refererred to as ``adiabatic''.  More
complicated models can also produce isocurvature perturbations, in
which density fluctuations in some species are initially exactly
compensated by density fluctuations in other species.  The data are
consistent with pure adiabatic and inconsistent with (at least the
simplest) pure isocurvature models.  Inflation also generically leads
to Gaussian, statistically isotropic perturbations whose statistical
properties are completely described only by a correlation function or,
in three-dimensional Fourier space, a power spectrum, $P(k)$.
Typically what is assumed in comparing CMB data with models is that
there is an initially adiabatic set of perturbations with power
spectrum parameterized by an amplitude and power--law spectral index,
$n$.

When the scale of a density perturbation is greater than that of the
Hubble length, the nature of the separate components (dark matter,
photons, baryons) is irrelevant.  However, because of their different
equations-of-state and different inter--species interactions, the small-scale 
dynamics of perturbation evolution is
different for the various components. When the Universe has aged
sufficiently that a wave of some given size is of a scale comparable to
the Hubble length (we casually say that the scale has 
``entered the horizon''), pressure and gravitational potential
gradients become important.  These gradients drive sound waves
in the multi-component plasma.  An important length scale for this
dynamical process is the {\em sound horizon}, the distance a sound 
wave could have travelled since
the Big Bang. Because the Universe has been dominated by radiation (with
sound speed $c/\sqrt3$) for most of its history, the sound horizon is
about ($1/\sqrt3$) of the (classical Big Bang) particle horizon.

First, consider waves entering the (sound) horizon around the time of
last scattering: these are the largest waves that could have formed a
coherent structure at this time. Indeed, by determining the
characteristic {\em angular} scales of the CMB fluctuation pattern,
and matching this to the physical scale of the sound horizon at last
scattering we can determine the {\em angular diameter distance} to the
last scattering suface, which is mostly dependent on the geometry of
the Universe: in a flat universe, angular and physical scales obey the
usual Euclidian formulae; in a closed (positively curved) universe,
geodesics converge and a given physical scale corresponds to a larger
angular scale (and hence smaller multipole $\ell$); conversely, in a
negatively curved Universe the same physical scale corresponds to a
smaller angular scale and largel $\ell$.

Consider now a wave that enters the horizon some time considerably
before Last Scattering, when the density of the Universe is still
dominated by radiation, and the Baryons are tightly-coupled to the
photons. Although the dark matter is pressureless, the dominant
radiation has pressure $p=\simeq\rho/3$, or somewhat less due to the
baryons.  Although the dark matter can continue to collapse, the
radiation rebounds when the pressure and density become sufficiently
high. Eventually, gravity may take over again and cause the perturbation
to collapse yet again, one or more times.  Larger and larger scales,
entering the horizon at later and later times, will thus experience
fewer and fewer collapse and rebound cycles. Moreover, because of the
effect of the baryons on the pressure, the strength of the rebound is
decreased as we increase the baryon density.  It is this cycle of
collapse and rebound that we see as peaks in the CMB power spectrum,
often called {\em acoustic peaks} after the acoustic waves responsible
for them. We thus use their heights to measure the relative
contributions of baryon and photons to the pressure, and their angular
scale to determine the geometry, as well as the history of the sound
speed in the baryon-photon plasma. (And of course other cosmological
parameters also affect the spectrum in yet other ways)

There are yet other physical effects that affect the power
spectrum. Although the photons and baryons are tightly bound to one
another via scattering, the coupling is not perfect. Hence, there is a
scale (known as the \emph{Silk} damping scale) below which the photons
can stream freely and wash out perturbations. This free-streaming damps 
perturbations on small scales.

\begin{figure}[htbp]
  \includegraphics[width=4in]{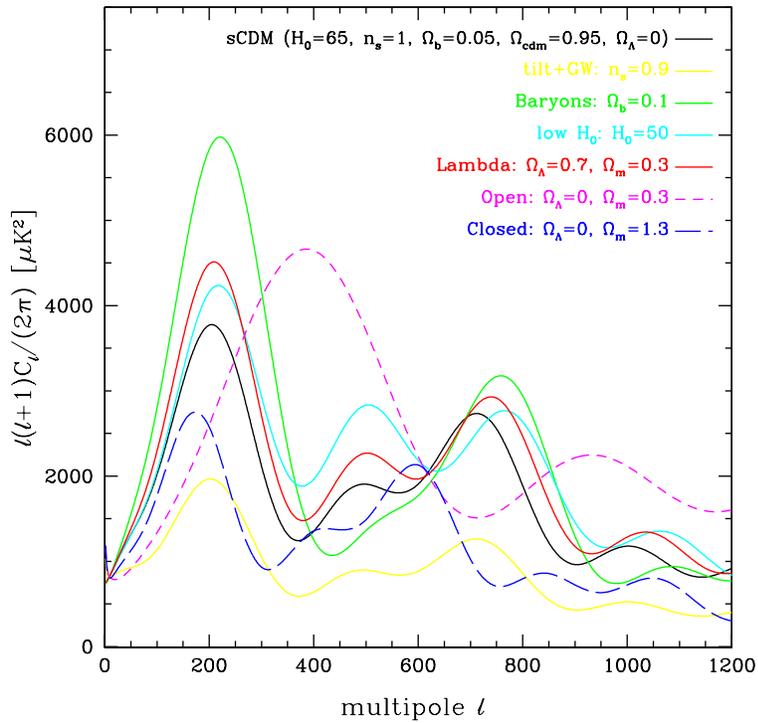}
  \caption{A sample of theoretical power spectra for various
    cosmological parameters, as marked.}
  \label{fig:theoryCl}
\end{figure}

All of these effects are included in codes~\cite{cmbfast,camb} which
solve the combined Boltzmann and linearized Einstein equations in an
expanding Universe.  These codes allow one to calculate the CMB
temperature power spectrum for a given model. A sample of spectra for
various input cosmological parameters is shown in
Figure~\ref{fig:theoryCl}.

The evolution of a single wavelength of perturbation is determined by a 
second order differential equation and thus has two independent solutions.
When the Hubble length is smaller than the perturbation wavelength (which
is the case at very early times, from soon after the perturbations creation
in the inflationary epoch) only one of these independent solutions
does not decay with time.  Thus all perturbations are ``squeezed'' into
the same state.  A result is that all acoustic oscillations of a 
given wavelength all have the same temporal phase.  This coherence
is important to achieving the multiple peak structures seen in
Figure~\ref{fig:theoryCl}.

The matter transport caused by the pressure and gravitational
potential gradients means there are {\em velocity} perturbations as
well.  Hence, the photons can scatter off of moving electrons, which
generates a net linear polarization of the photons~\cite{polprimer}.  
For the sound
waves we are considering, the velocities are greatest when the density
contrast is smallest, and vice versa: the velocity is out of phase
with the density--and hence the polarization signal is out of phase
with the temperature. Unfortunately, due to the relative inefficiency
of scattering off of the moving electrons, the polarization fraction
is only about 10\%, and the polarization spectra are
correspondingly suppressed.  

We discuss polarization induced by gravitational waves in 
subsection~\ref{motivationP}.

\section{Current Status of CMB Measurements}
\label{current}
%% to be written by AJ

\begin{figure}[htbp]
  \includegraphics[width=4in]{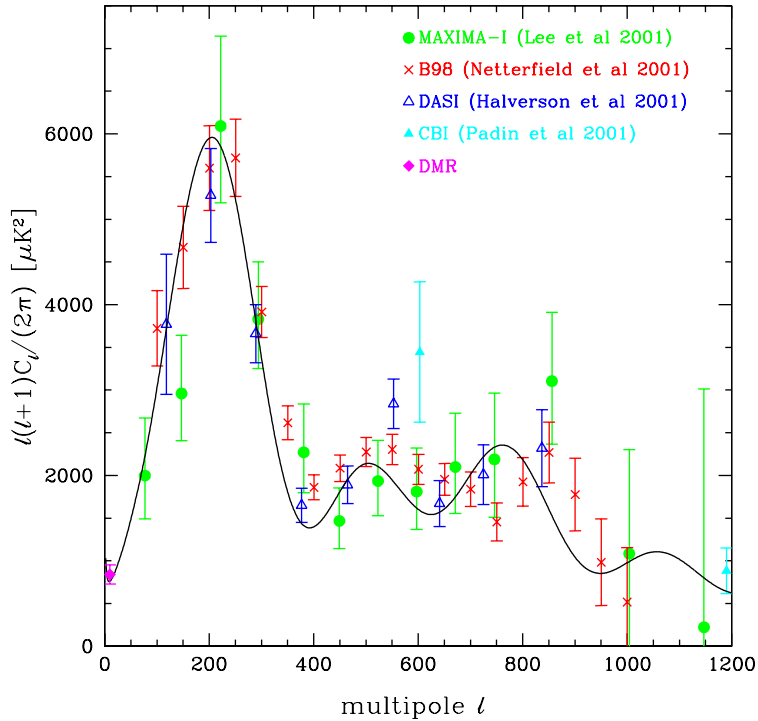}
  \caption{Recent measurements of the CMB power spectrum, from the
    experiments as listed and cited in the text. The smooth curve is a
    model chosen to fit an older subset of the data~\cite{maxiboom01},
    but remains a good fit to the current data.}
  \label{fig:Cl}
\end{figure}

In a real experimental setup, we measure the temperature (or intensity)
of radiation, and its polarization in various directions. Traditionally,
we measure components called $Q$ and $U$ giving two components
$45^\circ$ apart (there is also another polarization component, $V$
which is zero in cosmological situations). Thus, we start with noisy
measurements of $T$, $Q$ and $U$ on the sky, smeared by the instrumental
beam. If we can extract the underlying signal, we can then relate these
measurements to the power spectra predicted by theory. First, consider
the temperature signal. We start with the temperature pattern on the sky,
$\Delta T({\bf\hat x})/T=[T({\bf\hat x})-{\bar T}]/{\bar T}$, where
${\bar T}$ is the average temperature and ${\bf\hat x}$ is a unit
vector, and expand this in spherical harmonic multipoles:
\begin{equation}
  \label{eq:sphereharmonic}
  \frac{\Delta T}{T}({\bf\hat x}) = \sum_{\ell m} a^T_{\ell m} Y_{\ell m}({\bf\hat x})
\end{equation}
Under the assumptions of Gaussianity and an isotropic distribution on
the sky, we can treat the components $a^T_{\ell m}$ as if they were drawn
from a multivariate (but uncorrelated) Gaussian distribution with
variance
\begin{equation}
  \label{eq:Cldef}
  \langle a^T_{\ell m} a^T_{\ell' m'}\rangle = C_\ell
  \delta_{\ell,\ell'}\delta_{m,-m'}\; .
\end{equation}
Then, our task will be to determine $C_\ell$ from an actual noisy
realization of some part of the sky. In the class of inflationary
theories, these $C_\ell$ completely determine the statistics of the
temperature pattern, and are completely determined by the cosmological
parameters.

For polarization measurements, things are somewhat more complicated: we
must first relate the $Q$ and $U$ measurements to the ``gradient'' and
``curl'' fields on the sky ($a_{\ell m}^G$ and $a_{\ell m}^C$, using
so-called ``tensor spherical harmonics'' in place of the scalar $Y_{\ell
  m}$ of Eq.~\ref{eq:Cldef}) (which can only be done statistically in
the realistic case of noise, finite sky, and finite beam resolution),
and in turn relate these to the various power spectra:
\begin{equation}
  \label{eq:ClXX}
  C_\ell^{XX'}=\langle a_{\ell m}^X a_{\ell m}^{X'} \rangle,
\end{equation}
where $X,X'=T,G,C$. From parity considerations it can be shown that only
$C_\ell\equiv C_\ell^{TT}$, $C_\ell^{TG}$, $C_\ell^{GG}$ and
$C_\ell^{CC}$ are nonzero (except in the presence of parity-violating
physics). (All of this is a review of material first presented in
~\cite{KamKosSte,ZalSel97})

\subsection{Temperature and Polarization Power Spectrum Constraints}

In Figure~\ref{fig:Cl}, we show a selection of recent measurements of
the CMB temperature power spectrum, $C_\ell\equiv C_\ell^{TT}$. At the
lowest $\ell$ (corresponding to the largest angular scales) we show a
single point representing the measurement from the DMR instrument on the
COBE satellite~\cite{cobe}. Over $50<\ell<1300$ we show points from four
recent experiments with angular resolution of better than 20 arcmin.
MAXIMA~\cite{Lee01} and BOOMERANG~\cite{netterfield01} are balloon-borne
experiments using bolometers to measure the intensity of radiation;
CBI~\cite{CBI} and DASI~\cite{halverson00,pryke01} are ground-based
interferometers.  The recently-launched MAP satellite~\cite{MAP} promises
to surpass these measurements over most of this regime. MAP will span
the frequency range 30-90 GHz with an angular resolution of about $10'$
at 90 GHz.  The Planck Surveyor satellite~\cite{Planck} is due to be
launched in 2007 and will give sample-variance-limited measurements of
the temperature power spectrum out to $\ell\sim1500$ where the
information content becomes negligible due to damping effects. Planck
covers a frequency regime of 30-850 GHz, with an angular resolution of
$5'$ at the highest frequencies.

\begin{figure}[htbp]
\includegraphics[width=1\columnwidth]{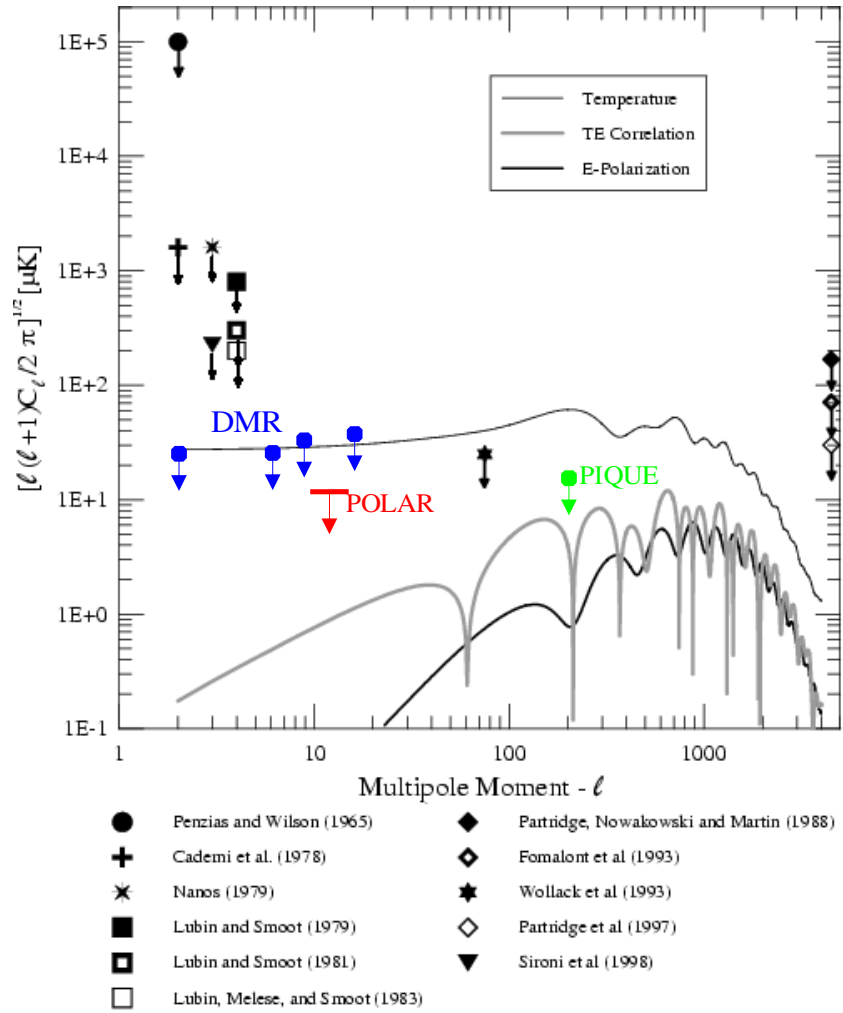}
  \caption{Current limits on the polarization of the CMB. From
    Ref.~\cite{StaggsGundrvw}, except for more recent points from
    POLAR~\cite{POLAR01}, PIQUE~\cite{PIQUE01} and unpublished limits from
    DMR~\cite{DMRpol}.}
  \label{fig:pol}
\end{figure}

In contrast, we are only now reaching the requisite sensitivity to
detect polarization; upper limits on the polarization power spectrum are
shown in Figure~\ref{fig:pol}; they are now within an order of
magnitude of the predicted RMS polarization.

Although measurements of the CMB are providing remarkable
quantitative constraints on cosmological parameters, we begin by
emphasizing the following qualitative interpretations of the data:
\begin{itemize}
\item the initial perturbations were adiabatic with a 
nearly scale--invariant power spectrum,
\item the mean spatial curvature is near zero,
\item the acoustic oscillations are highly coherent.
\end{itemize}
These are all predictions of inflation.  Whether they are unique
to inflation is a matter of debate.  Scale--invariant
perturbations can be created without inflation and flat Universes
are perhaps preferred by quantum--tunneling events.  The
theoretical prejudice for these two items actually precedes the
first papers on inflation.  It is the third item, the coherence of
the acoustic oscillations, which makes the strongest case we have
for an alteration of the causal structure of the space--time in
some early epoch, as discussed in~\cite{coherence1}.

In brief, the coherence can result from a "non--dynamical" stage
of perturbation evolution---when the wavelength of a given
perturbation is larger than the Hubble radius.  Thus the challenge
is to create perturbations which are, for some period of time,
larger than the Hubble radius.  One cannot do this causally in a
classical big--bang model without a period of inflation.  This
statement is almost definitional, since inflation is an
acceleration of the scale factor and acceleration of the scale
factor is what is required for the Hubble radius to grow more
slowly than the scale factor.  Causal analyses become more
complicated in a brane--world picture, and 
there are scenarios in which super--Hubble radius
perturbations are created without any acceleration of the scale
factor~\cite{ekpyrotic}.

Quantitative cosmological constraints from current CMB
measurements are also of great interest.  As mentioned
earlier what these constraints assume from inflation is
an initially power--law power spectrum of adiabatic fluctuations.
One of the best--measured parameters from CMB experiments is
$\Omega_{\rm tot}$, which is equal to unity when the mean
spatial curvature is zero.  The results from the experiments
are quite clear.  From DASI, Boomerang and Maxima,
$\Omega_{\rm tot}=1.01 \pm .08, 0.97 \pm .10, 1.0 \pm .14$ respectively.
Combining all these experiments and others~\cite{wang01} results
in $\Omega_{\rm tot} = 1.0^{+.06}_{-.05}$.  
These results are significant for
inflation since a non--zero curvature would make inflation
a much less attractive early Universe scenario.   

%% constraints on n
The power--law spectral index for the initial power spectrum of
scalar perturbations, $n_s$, is also well--determined from the data.
Once again from DASI, Boomerang and Maxima we have 
$n_s = 1.04 \pm .06, 1.02 \pm .06, 1.08 \pm .10$ respectively.
The value $n_s = 1$ is scale--invariant since for that value
the contribution to the density variance from each decade of wavenumber
is constant.  Inflationary models produce values of $n_s$ in the
range 0.8-1.2.  Refined determinations of $n_s$ will thus not
by themselves test inflation, but will rule out particular classes of
inflationary models.

%% Omega_Lambda, h and age
The stronger one's model assumptions, the stronger the conclusions one
can draw from CMB data.  An interesting assumption, motivated by
simplicity and the success of inflationary models, is the assumption
that $\Omega_{\rm tot}=1$.  With the mean curvature thus fixed, one
can then use the peak locations (combined with assumptions about the
Hubble constant) to determine $\Omega_\Lambda$~\cite{kamionkowski00,
knox00} (or more generally some combination of the dark energy density
and equation-of-state parameter, $w$).  For example, if $h > 0.55$
\cite{knox01} report $\Omega_\Lambda > 0.4$ at 95\% confidence.  This
gives an argument for the existence of dark energy which is independent
of supernovae and constraints on the dark matter density.  

%% Weakening assumptions:   effect of gravity waves, isocurvature modes
Of course, the weaker one's model assumptions the less one can say, and
this model--dependence is important to keep in mind.  The model space
can be expanded in several ways.  Allowing for tensor perturbations
opens up $n_s$ to be in the``$2\sigma$'' range 0.89-1.49 according to
\cite{efstathiou01} and similar result in~\cite{wang01}.  As a more
extreme example, allowing for arbitrary admixtures of various isocurvature
modes together with the adiabatic mode makes most parameter constraints
from the CMB large enough to be useless \cite{trotta01}.

%% bringing in extra constraints (e.g., Ly-alpha forest).  Importance
%% of extending dynamic range for constraints on $P(k)$.
The model--dependence of the CMB parameter constraints underscores
the importance of other cosmological measurements such as galaxy
redshift surveys, quasar spectrum studies of the Lyman-$\alpha$ forest
and supernovae observations.  These observations can either serve
to support the modeling assumptions or further tighten
parameter uncertainties.  For example, adding in galaxy redshift
survey information and BBN constraints on the baryon density
shrinks the $n_s$ range cited above~\cite{efstathiou01} by a factor
of 3.

For probing inflation, we want the best information we can get
on the primordial power spectrum, $P(k)$.  In particular, it
would be very helpful to see some departure from an exact power--law.
To see any break we want to have as long a lever arm as possible
and thus we want to complement CMB measurements with determinations
of $P(k)$ on small scales.  The best way to do this, at the moment,
is with Lyman--$\alpha$ forest observations.~ \cite{croft00}  For the
first attempts see Refs.~\cite{hannestad01,zaldarriaga01}.

\section{Motivation for Further Measurements}
\label{motivation}

\begin{figure}[htbp]
\includegraphics[width=4in]{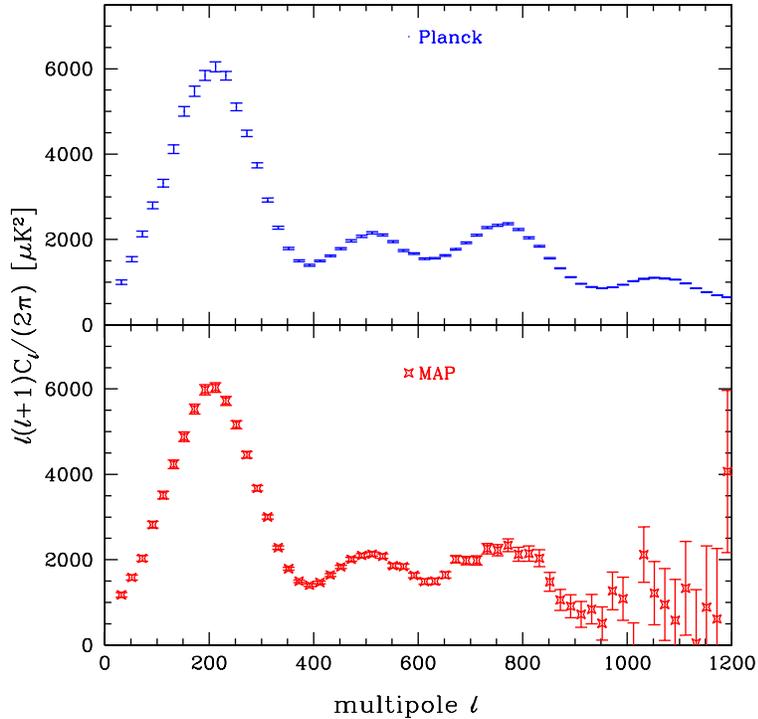}
   \caption{Projected error bars for the power spectrum from Map
     (bottom) and Planck (Top). We have binned the spectra with a width
     $\delta\ell=20$.
}
   \label{fig:mapplanck}
\end{figure}

Two satellite missions that will greatly improve upon current measurements of the
CMB are {\it MAP} and the {\it Planck Surveyor}.  Simulated 
power--spectrum determinations for these missions are shown in
Figure~\ref{fig:mapplanck}.  As has been detailed elsewhere, these measurements will 
enable us to determine cosmological parameters to unprecedented precision
\cite{forecast}, to reconstruct the primordial density perturbation spectrum,
$P(k)$~\cite{wang99}, and discriminate among inflationary models (see Figure~\ref{lek-infl-fig1}).

Here we provide motivation for measurements with even higher sensitivity than
those to be made by {\it MAP} and {\it Planck}.  This higher sensitivity is necessary for
measuring the polarization of the CMB and determining higher--order temperature correlations.
For a review of future experiments designed to measure the temperature and
polarization of the CMB, see the proceedings of working group E6.1 in this
volume~\cite{scgkt01}.  
\subsection{Temperature}
\label{motivationT}

The {\it Planck Surveyor}, if it performs to specifications,
will be practically the final word on the study of {\it primary}
CMB temperature anisotropy.  For $l \la 1500$, all the $C_l$'s will
be determined to nearly the sampling variance limit; i.e., further
reduction in the measurement noise will not reduce the $C_l$ uncertainty.

Further refinement of our knowledge of CMB temperature anisotropies is
nevertheless well--motivated.  With higher angular resolution and
higher sensitivity maps one can study secondary effects in the CMB
anisotropy, created much more recently than last--scattering.  These
secondary effects are created both gravitationally (via lensing and
the ``Integrated Sachs--Wolfe'' (ISW) effect) and by Thomson
scattering off of electrons in the post--reionization inter--galactic
medium.  Of particular interest for this report are the lensing and
ISW effects because these can be used to break the degeneracy between
$\Omega_\Lambda$ and $w$ which exists in constraints from the primary
CMB data alone.

\begin{figure}[htbp]
\includegraphics[width=5in]{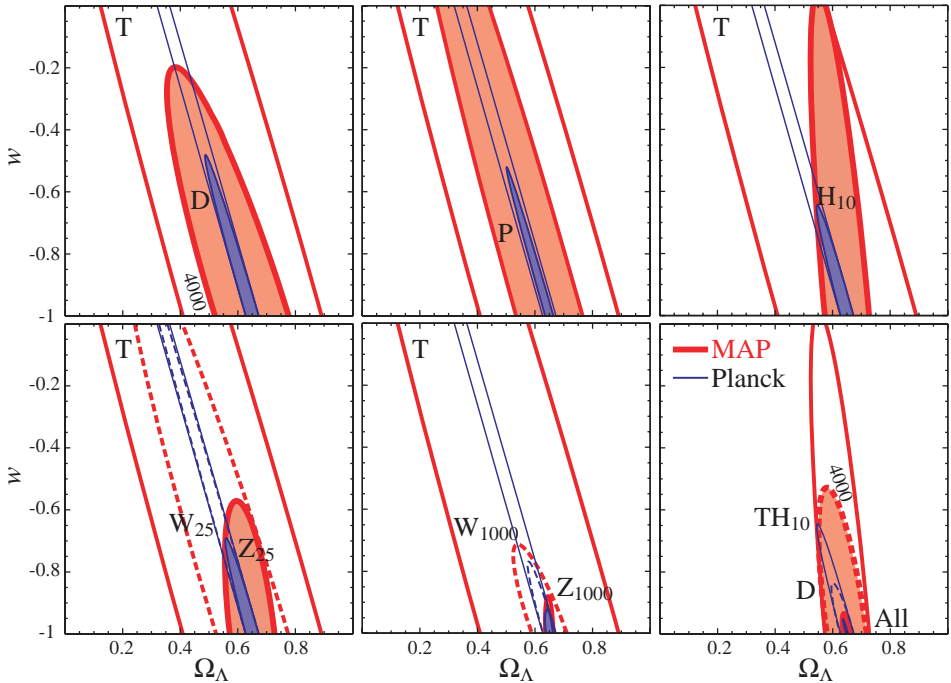}
   \caption[]{
Improvement on the MAP (thick) and Planck (thin) temperature ($T$)
determination of the dark energy
equation of state and density.  Clockwise from the top left: 
addition of CMB deflection angles ($D$);
polarization ($P$); 10\% Hubble constant measurements ($H$); 25 deg$^2$ cosmic shear survey 
with ($Z$ solid) and without ($W$ dashed) tomography; same but for 1000 deg$^2$; Hubble constant ($TH$),
plus deflections ($D$), plus 
a 1000 deg$^2$ lensing survey (All)
(from \cite{hu01}).}
   \label{lek-motivation-fig1}
\end{figure}

Hu~\cite{hu01} considers complementing Planck with a CMB survey (called
the ``D'' survey) covering one tenth of the sky with 1' resolution
(fwhm) and 10~$\mu$K per pixel errors.  Such a map would be a
probe of gravitational lensing and ISW effects in the CMB.
The correlations between these two effects contribute to the
three--point correlation function and depend on the amount of dark
energy and its equation--of--state.  Forecasted constraints on
$\Omega_\Lambda$ and $w$ from MAP and Planck, complemented by the D survey,
are shown in the top left panel of Fig.~\ref{lek-motivation-fig1}.
Unfortunately, the constraints on $w$ are not very strong.  

Since the lensing and ISW are late--time effects, associated with the
more--local Universe, there is great value to be had from
cross--correlating with other observables.  Of particular value are
galaxy lensinging surveys with source galaxies broken up into coarse
redshift bands (Z).  Indeed, the impact of such a survey in the
$\Omega_\Lambda$--$w$ plane is much greater than for the D survey, as
one can see in the lower panels of Fig.~\ref{lek-motivation-fig1}.  
Combining Planck with a 1000 sq. deg. Z survey leads to an error
on $w$ of .05.  Adding the D survey to Planck and Z reduces the error 
to 0.03.  See \cite{hu01} for details.  Also see \cite{verde01b}.

%% Hu

%% Comment on foreground effects
Astrophysical foregrounds have proven to be fairly benign for
measurement of the CMB temperature $C_l$ over the important range of $l$ for
primary CMB anisotropy.  Whether this happy situation will extend
to higher angular scales and to the higher sensitivity required for
studying non--Gaussian properties of secondary anisotropies is not
yet clear.  Studying the non--Gaussian properties of available foreground
maps would be a good first step towards incorporating these concerns
into error forecasting.

\subsection{Polarization}
\label{motivationP}

In section ~\ref{acoustic} we considered polarization due to electrons moving with
the fluid as it responds to the pressure and gravitational potential gradients.
The polarization pattern that results is
``curl-free''; that is, it can be represented as the 2-d divergence
(gradient) of some scalar on the sky.  This is easy to understand if we
heuristically identify the polarization pattern with the flow of the
plasma.  In linear perturbation theory this flow is a potential flow;
any vorticity is damped by expansion and not driven to grow by gravity.

Gravitational waves at last--scattering can produce a ``curl''
pattern, since gravity waves can produce non--potential flows in the plasma.
For full-sky maps with high sensitivity and resolution, the ``gradient''
and ``curl'' components (also known as ``Electric'' and ``Magnetic'' or
$E$ and $B$ from the obvious analogy to electromagnetic fields) can be
separated completely; in more realistic situations statistical
techniques are necessary.

Unfortunately, gravitational waves from inflation are not the only way
to create polarization patterns with non--zero curl.  Astrophysical
foregrounds, such as emission from dust in our own galaxy and
extragalactic point sources, will also produce some curl.  These can
be subtracted to some degree with the aid of multi--frequency
measurements since they are spectrally distinct.  Since their spectra
aren't perfectly known, however, there will always be some residual
uncertainty in the subtraction.  More insidious is the production of a
curl pattern by non--linear evolution.  Specifically, the deflection
of light rays through latter-day large-scale structures will generate
a curl pattern\cite{hu01,ZalSel98,LewChaTur01b}.  This contribution can
also be distinguished from the gravitational wave contribution via the
shape of its power spectrum.  However, the uncertainty in this
subtraction will put a fundamental limit on the sensitivity of any
polarization measurement to gravitational waves.  This limit has not
yet been calculated exactly but it's approximately $r=10^{-3}$
\cite{LewChaTur01b} where $r$ is the ratio of gravitational wave to
density perturbation contributions to the temperature quadrupole
variance, $C_2$.  Translated into an energy scale of inflation, if
this energy scale is below about $V^{1/4}= 5 \times 10^{15}$ GeV then
$r < 10^{-3}$ and we will never detect the influence of the gravity
waves.

Detection of a curl pattern with the right photon spectrum and power spectrum
would be very interesting.  The amplitude of this pattern is proportional
to the energy--scale of inflation.  There is no other way of getting this
information.  And it would be more evidence of
inflation as the generator of perturbations.  Possibly this evidence
would constitute a ``smoking gun'' but there is no proof of inflation's
uniqueness in this regard.  Interestingly, one possible alternative for the 
generator of density perturbations, 
the `ekpyrotic Universe', produces a gravity-wave background with a very small 
amplitude and bluer spectrum~\cite{ekpyrotic}.  Other mechanisms for
producing gravity waves (such as first order phase transitions) 
result in redder power spectra.  

Whereas the Planck Surveyor will recover the entire primary temperature
anisotropy, neither Planck nor any other currently planned and funded
missions will achieve the same for polarization. Only a small
fraction of the ``E-mode'' multipole moments from 2 to 2000 will be
measured with signal--to--noise ratios greater than one.  
For the ``B-mode,'' if it comes from
inflationary gravity waves, the situation is even less certain, since
the overall amplitude of the signal depends on the energy scale of
Inflation. 

As our knowledge of CMB polarization increases --- which it should do
very rapidly over the next very few years, as we first detect it
(expected in 2001-2002) and then measure the polarization spectra with
some precision, we will begin to get a handle on the orders of magnitude of
the instrumental and astrophysical problems. With this knowledge, we
will be able to consider a more community-wide effort to measure
polarization with the sensitivity necessary for an exploration of the
B-mode signal (if present, of course). The time is ripe today to
continue development of appropriate high-sensitivity technologies.  What is
needed is either radically new technologies for individual detectors or
increasing the number of detectors from the $\sim100$ aboard Planck by
more than an order of magnitude. Whether such an instruments needs to be
in space is still unclear and will depend in part on the magnitude of the foreground
problem, and thus on the range of frequencies we will be required to cover in order 
to cleanly extract the cosmological signal.
For more details on the hardware
aspects of this problem, see the companion report of section E6.

\section{Analysis Challenges}
\label{analysis}

As the amount of CMB data has increased over the last decade, our
understanding of the optimal analysis techniques has progressed
alongside (for a fuller description, see, e.g., \cite{BCJK99}). The
analysis pipeline can be seen as several subsequent steps of data
compression\cite{BJK00}: from the raw data taken by the instrument, to a
cleaned and calibrated timestream of data; from the timestream to a map
on the sky (or, in the case of interferometers, to a set of
visibilities); from the map (or visibilities) to the CMB power spectrum;
and finally from the power spectrum to the cosmological parameters. 
%Of
%course, in addition to greatly reducing the amount that must be treated
%at each step, the intermediate products (the timestream, the map, and
%the spectrum) are all intersesting in their own right, both as
%diagnostics of the underlying assumptions going into the procedure, and
%as scientific results in their own right.

In the standard analyis techniques used today, we perform each of this
steps using a Bayesian or maximum-likelihood formalism, usually under
the assumption that the distributions of both the instrumental noise and
the underlying cosmological signal can be treated as multivariate
Gaussians, with covariance matrices given, respectively, by the
instrumental noise characteristics (which must themselves be determined
from the data\cite{FerJafMNRAS00,Dore01}), and the CMB power spectrum. This in
turn means that the likelihood calculations require the manipulation of
these matrices. Unfortunately, the manipulations required, at least for
the brute-force solution to the likelihood equations, scale as the cube
of the number of pixels. For data sets of even moderate size ($\sim
10,000$ pixels) this becomes prohibitive on an individual workstation
(circa 2001); Borrill\cite{MADCAP} has written the MADCAP package to
perform the mapmaking and spectrum-estimation steps using standard
libraries available on parallel supercomputers.

We note that the vast majority of analysis time
is still spent understanding the systematic problems of any individual
experiment (and even different incarnations of the same experiment); for
examples of the work required, see\cite{stompor01Maps,netterfield01}.

These quantitatively and qualitatively new data will bring with them 
new problems in their analysis. The first and most obvious is simply the
amount of data. As stated above, the brute-force solution to the
likelihood problem scales as the cube of the number of pixels. For the
megapixel datasets of MAP and Planck, this will likely be impossible
even with expected increases in processing power over the required
timespan. Several new methods and ideas have already been explored.
Here, we will discuss recent work in the most computationally-intensive
phase of the analysis, the determination of the power spectrum.

The first possibility takes note of the fact that one reason for the
$N^3$ scaling is that the instrumental noise and the CMB fluctuations
have very different natural bases: the CMB fluctuations are expressed
most naturally as spherical harmonics, whereas the instrumental noise is
expressed in the timestream, or, after the mapmaking manipulations, in
the pixel basis. In \cite{OSH99}, the authors show, for the special case
of the expected performance of the MAP satellite, the instrumental noise
can also approximately be expressed in the spherical-harmonic bases, and
this approximation used as the basis for iterative schemes for the
required matrix manipulations.

Another possibility uses the folk wisdom that high-$\ell$ information
can be gathered separately from different parts of the sky, without
performing the full analysis, whereas low-$\ell$ information only
requires a smoothed version of the full data. In \cite{DoreKnox01}, the
authors use this as the basis of a hierarchical decomposition of the
dataset, iterating toward an approximation to the likelihood by
combining the results for maps at different resolution levels and areas
of the sky.

%Pseudo-$C_l$

%Ring data?

Yet another possible solution abandons the Bayesian approach entirely
and tries to find frequentist statistics as estimators for the power
spectrum\cite{Hivon01}. That is, we first select some ``natural''
estimator for the power spectrum, such as the square of the windowed
Fourier components of the noisy map, which is then modified by
appropriate multiplicative and additive filters, determined so that the 
estimator is suitably unbiased with appropriately small variance.
Finally, there has been some work (used in \cite{netterfield01})
using these frequentist statistics within the likelihood formalism, as
approximations to the shape of the likelihood in the ``Bayesian'' sense
(as functions of the $C_l$ for fixed data).

All of the discussion so far has been directed toward the analysis of
temperature data; polarization data presents its own set of
problems. The most obvious is simply size, yet again: instead of just
the temperature at each pixel, we now have two additional numbers
describing the polarization: to analyze a map of the same size by
brute-force techniques will take $\sim3^3=27$ times as long. Moreover,
the very low signal-to-noise of the polarized data will require far
greater attention to systematic problems (and foregrounds) than is
currently required by temperature data; only with real data in hand
will we be able to explore these problems in any detail.

\section{SZ Surveys}
\label{sz_v4}
As discussed in Section~\ref{current} accurate measurements on the CMB
will set tight constraints on the relative densities of baryonic
matter, dark matter and dark energy {\em at the surface of last
scattering}.  In addition, one particular combination of these
densities, the curvature, and the dark energy equation of state
parameter, $w$, will be well--determined---that combination which
fixes the ratio of angular--diameter distance to last-scattering
surface to the sound horizon at last scattering.

However, as shown in Fig.~\ref{lek-motivation-fig1}, CMB measurements
cannot, by themselves, separately determine the dark energy
density and $w$.  In particular, they are highly unlikely
to distinguish a cosmological constant ($w=-1$) from any of
the currently viable alternatives.  For example, the dark
energy may be the result of a slowly evolving
scalar field, with $w> -1$ and time--varying~\cite{ws98, wco99}.

Measurements of large scale structure offer good prospects for probing
$w$. Dark energy manifests itself as a smooth component that
will inhibit the gravitational collapse of structure once the dark energy
density begins to exceed the density of gravitating matter.  The value
of $w$ also affects the comoving volume in solid angle $d\Omega$ from
redshift $z$ to redshift $z+dz$, and therefore how comoving densities
are translated into observables.

Galaxy clusters are the largest virialized objects in the universe and are
believed to have formed relatively late (at $z=1$--3). Clusters are
important cosmological probes for several reasons. First they represent
the accumulation of matter from quite large regions of the universe and so
one may assume that their contents are representative of the universe as a
whole. Most of the baryonic matter in clusters is contained in hot gas
trapped in the cluster potential well which can be detected via the
Sunyaev-Zel'dovich effect (see below) or from X-ray measurements of
bremsstrahlung emission. The total mass of the cluster can be deduced from
weak gravitational lensing if such data are available, or from the spatial
distribution of the gas, with an assumption that the gas is in hydrostatic
equilibrium. In this way clusters can be used to constrain
$\Omega_b/\Omega_m$~\cite{2001ApJ...552....2G} where $\Omega_b$ and
$\Omega_m$ are the density parameters of baryonic matter and all matter,
respectively. Second, the details of cluster formation can be modeled in a
straightforward way since they depend primarily on the details of the
gravitational collapse of structure, and not on more ``messy'' processes
such as gas dynamics and energy injection from star formation (although
see Section~\ref{sec-sz-smod}). As a result, the masses of clusters and
their distribution in redshift is primarily sensitive to purely
cosmological factors such as:
\begin{itemize}
\item The initial power spectrum of matter fluctuations, characterized by
$\sigma_8$, the rms fluctuation in mass in a region with radius
8$h^{-1}$\,Mpc, and $n$, the spectral index of density fluctuations on the
scale of clusters.
\item The rate of growth of structure via gravitational collapse.
\item The geometry of the universe, since this determines the size of a volume
element at a given redshift.
\end{itemize}

As discussed below, the Sunyaev-Zel'dovich effect offers excellent
prospects for using galaxy clusters in this way.

\subsection{The Sunyaev-Zel'dovich Effect}
The Sunyaev Zel'dovich (SZ) effect~\cite{sz72} is the result of the
interaction of the CMB with ionized gas along a line-of-sight to the
surface of last-scattering. Compton-scattering of CMB photons by the much
hotter electrons in the gas causes a distortion, $\Delta I_{\rm th}$ to
the intensity, $I_{\rm CMB}$, which in the non-relativistic limit is given
by:
\begin{equation}
 \frac{\Delta I_{\rm th}}{I_{\rm CMB}} = \frac{x e^x}{(e^x-1)} \left[
x\coth\frac{x}{2} -4\right] \times y_{\rm th} \mbox{\hspace*{0.5cm} where
\hspace*{0.5cm} } y_{\rm th}= \sigma_T \int n_e \frac{kT_{\rm e}}{m_{\rm
e} c^2} {\rm d}l
\label{sec-sz-eq1}
\end{equation}
with $x=h\nu/kT_{\rm CMB}$.  The quantity $y_{\rm th}$ is proportional to
the pressure of the gas integrated along the line-of-sight to the last
scattering surface and depends on $T_{\rm e}$, the temperature of the gas,
$\sigma_T$, the Thompson cross-section, and $n_e$, the electron density.
The hottest gas is located in cluster potential wells and can be as hot as
15\,keV, and consequently clusters will dominate the SZ signal. However,
the CMB acts as a uniform back-light to {\em all} of the hot gas in the
universe, including the cooler, less-dense gas with temperature 8--800 eV
which is predicted from simulations of large scale structure formation to
exist as a filamentary structure between clusters.~\cite{cen99}

\begin{figure}
\includegraphics[width=4in]{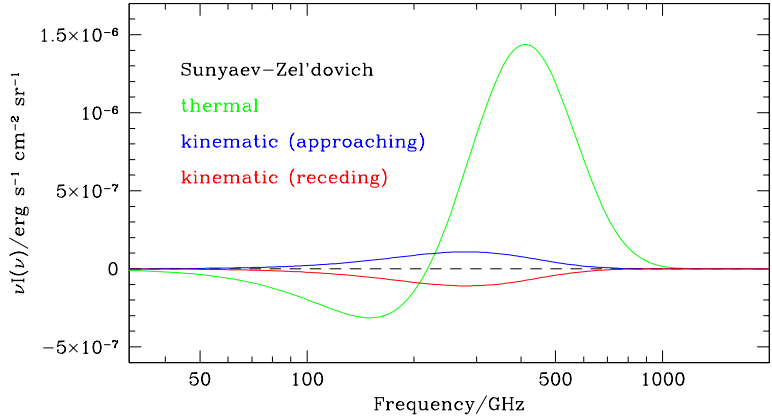}
   \caption[]{Brightness of the SZ effect as a function of frequency. The
solid line is the thermal component (in the non-relativistic limit), the
two dotted lines show the kinetic component. The sign of the kinetic
component depends on the direction of the cluster peculiar velocity
relative to the observer. The assumed parameters are $\tau=1$\%, $T_e =
5$~keV, and $v_{\rm pec}$=1000~km/s.}
   \label{sec-sz-fig1}
\end{figure}

The distortion characterized by $y_{\rm th}$ is known as the thermal SZ
effect because the amplitude is related to the thermal motions of the
electrons in the clusters.  The thermal SZ effect has a unique spectral
shape (see Fig.~\ref{sec-sz-fig1}) causing a rich cluster to appear as a
``hole'' in the CMB at low frequencies, but as an increase in the CMB
intensity at frequencies above $\sim 217$\,GHz. Eq.~\ref{sec-sz-eq1} is
valid only in the non-relativistic limit, so for most rich clusters,
corrections to the spectrum at the level of a few percent are
necessary~\cite{nsi00,reph95}. This causes the frequency of the null of
the thermal effect, and the overall shape of the spectrum, to be weakly
dependent on the gas temperature.

In addition to the thermal effect, there is a second component -- the
kinematic SZ effect -- which is the result of the bulk motion of the
plasma in the rest frame of the CMB. In the non-relativistic limit, the
change in brightness is given by:
\begin{equation}
\frac{\Delta I_{\rm kin}}{I_{\rm CMB}} = \frac{x e^x}{(e^x-1)} \times
y_{\rm kin} \mbox{\hspace*{0.5cm} where \hspace*{0.5cm}} y_{\rm kin} =
\sigma_T \int n_e \frac{{\bf v}_{\rm pec}\cdot {\rm d}{\bf l}}{c} = \tau
\frac{v_{\rm pec}}{c}
\label{sec-sz-eq2}
\end{equation}
Here $v_{\rm pec}$ is the mean radial component of the peculiar velocity
of the cluster plasma, ${\bf v}_{\rm pec}$. The optical depth, $\tau$, of
a rich cluster is typically 1\%. The spectral profile of the kinematic SZ
effect is also shown in Figure~\ref{sec-sz-fig1}.  The kinematic effect
has yet to be detected, but since cluster peculiar velocities are expected
to be no more than a few hundred km/s~\cite{1994ApJ...430L..13B,
1996MNRAS.282..384M} the kinematic effect is likely to be at least an
order of magnitude fainter than the thermal effect.

The expressions in Eqs.~\ref{sec-sz-eq1} and~\ref{sec-sz-eq2} allow the
determination of the SZ effect along a given line of sight through a
cluster, This quantity has {\em no explicit redshift dependence} because
the effect is a scattering process and both $\Delta I$ and $I_{\rm CMB}$
scale in the same way with redshift. Consequently, the amplitude and
spectral shape of the SZ effect are {\em independent} of the distance to
the cluster. Of course, the redshift does affect the total SZ flux from a
cluster. Assuming that a cluster is approximately isothermal, the
integrated flux from a cluster at redshift $z$ is:
\begin{equation}
S = \frac{i_00(\nu)}{d_A^2(z)} \times \frac{kT_e}{m_e c^2} \frac{\sigma_T
f_{\rm ICM}}{\mu_e m_p} \times M_v
\label{sec-sz-eq3}
\end{equation}
where:
\begin{equation}
i_0(\nu)=2kT_{\rm CMB} \left(\frac{kT_{\rm CMB}}{hc}\right)^2 \frac{x^4
e^x}{(e^x-1)^2} \left[x\coth\frac{x}{2}-4\right]
\label{sec-sz-eq4}
\end{equation}
The quantity $d_A(z)$ is the angular diameter distance to the cluster,
$M_v$ is the mass of the cluster contained in the virial radius, $\mu_e=1.15$
is the mean molecular weight per electron, $f_{\rm ICM}$ is the fraction
of the mass contained in the intracluster medium (ICM) and $m_p$ is the
proton mass. Various forms of Eq.~\ref{sec-sz-eq3} can be found
in Refs.~\cite{bbbo96,hc01,vhs01}.

From Eq.~\ref{sec-sz-eq3}, it can be seen that although there is a
dependence of total flux on redshift, it is much weaker than for other
sources of emission such as X-ray flux which decreases by an extra factor
of $1/(1+z)^4$ as the redshift increases. This makes the SZ effect a prime
technique for detecting clusters of galaxies out to the epoch of cluster
formation.

\subsection{Prospects for SZ Measurements}
\label{sec-sz-spros}
\begin{figure}
   \includegraphics[width=2.5in]{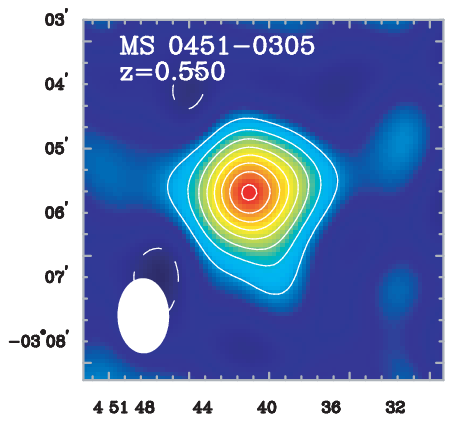} \hspace*{0.5cm}
   \includegraphics[height=2in]{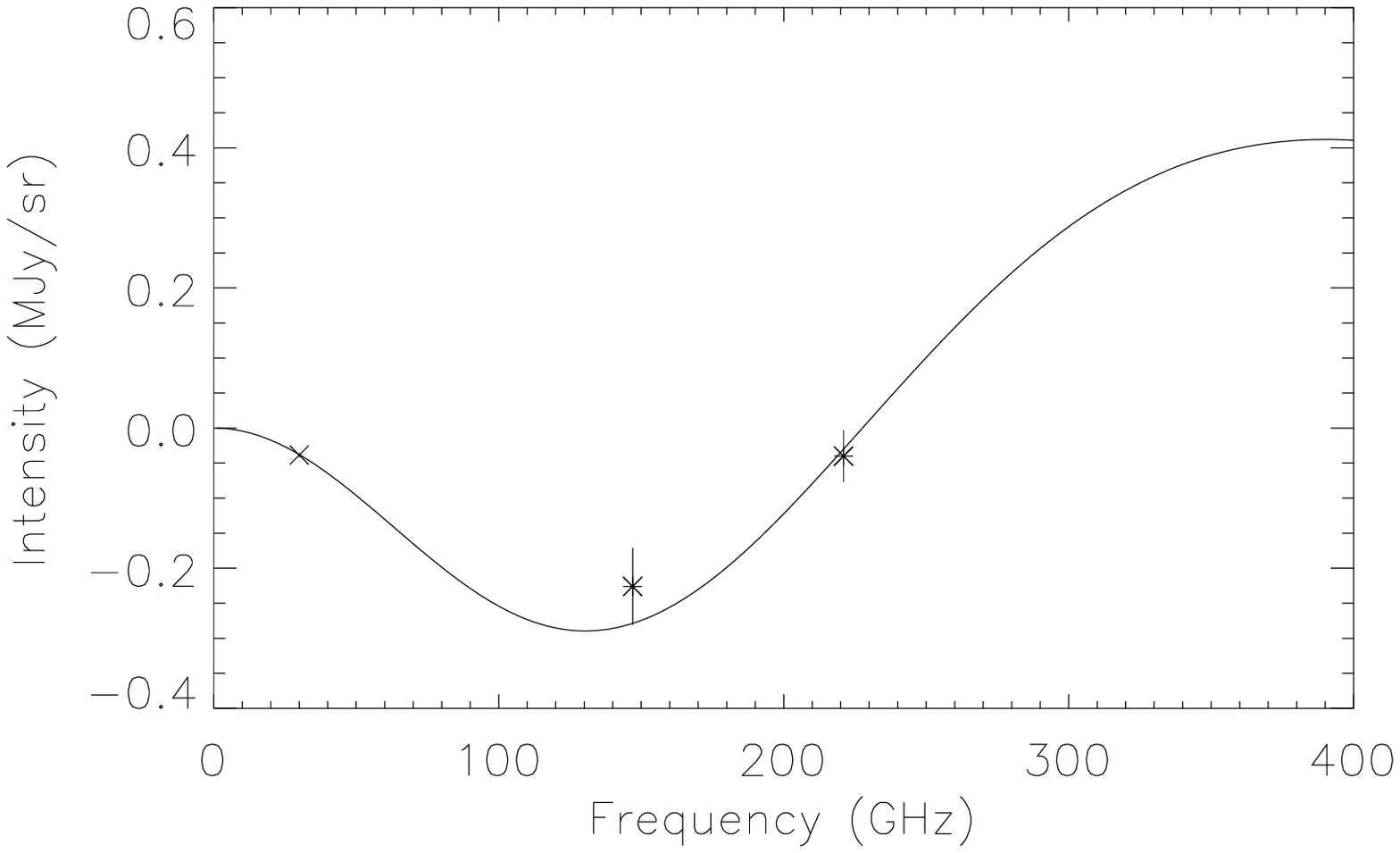}
   \caption[]{Left: Map of the SZ decrement towards the high redshift cluster MS0451
made using the BIMA array (from~\cite{rmc00}, courtesy J. Carlstrom.).
Right: Measurement of the spectrum of MS0451 using the flux measured with
the BIMA array (cross) and the SuZIE experiment (stars)~\cite{bc01}.}
   \label{sec-sz-fig4}
\end{figure}

For a comprehensive review of the current status and
future prospects of SZ measurements, the reader is referred to the
proceedings of E6.1 in this volume~\cite{scgkt01}, and also to
\cite{carl00}. The first detections of the SZ effect required many years
of pioneering work~\cite{1990cmwb.book...77B} but in the
past 5 years, detections of the thermal SZ effect in rich clusters
identified from X-ray surveys have become routine.  SZ measurements have
been used to obtain estimates of the Hubble constant, of the gas mass
fraction in clusters and of the peculiar velocities of galaxy clusters.
Fig.~\ref{sec-sz-fig4} shows the type of data that can now be routinely
obtained with existing instruments.

\begin{figure}
   \includegraphics[width=3in]{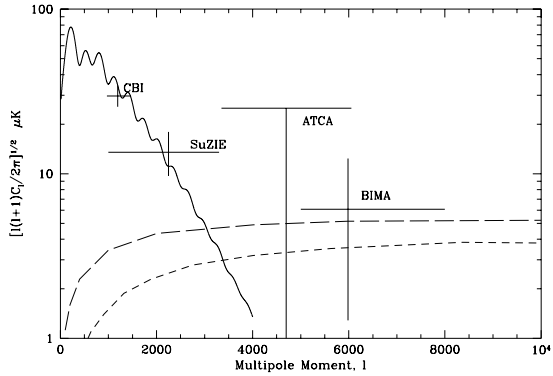}
   \caption[]{Current limits on the SZ power
     spectrum.~\cite{CBI,cb01,daw01,sub00} The solid line is the best-fit
CMB power spectrum to the CBI point (generated with CMBFAST~\cite{sel96}).
The two dashed lines are two models for the power spectrum of the thermal
SZ effect that reflect different assumptions regarding the evolution of
the cluster gas~\cite{1999mifo.conf..199H, 2001ApJ...549..681S} (see
Section~\ref{sec-sz-smod}). }
   \label{sec-sz-fig5}
\end{figure}

Blind searches with existing SZ instruments for clusters not previously
identified in optical or X-ray surveys have either been unsuccessful or
have produced ambiguous results which have not been verified elsewhere.
This is because existing instruments, although excellent for investigation
of known clusters, lack the sensitivity and sky coverage required and the
observations shown in Fig.~\ref{sec-sz-fig4} typically take many hours.
Fig.~\ref{sec-sz-fig5}, which shows current limits on the SZ power
spectrum give an indication of the gap between the current and the ideal
experiment. A new set of experiments has been proposed with emphasis on
large areas of sky and multi-frequency observations to separate SZ from
intrinsic CMB.  These experiments are reviewed in detail in the E6.1
proceedings in this volume~\cite{scgkt01} and can be divided into the
following catagories:
\begin{itemize}
\item {\bf Deep surveys} of about 10 sq deg. with radio interferometers should
achieve a limiting mass of $10^{14}$~M$_\odot$~\cite{carl98, kjs01} and
will obtain high resolution maps of individual clusters.
\item {\bf Medium-deep surveys} of about 100-4000 sq deg. using bolometer arrays on
6-m class telescopes expect to achieve a limiting mass of
$2$--$4\times10^{14}$~M$_\odot$~\cite{hhm01, vhs01}.
\item {\bf Shallow Surveys} The all-sky survey from the Planck Surveyor will achieve a mass
limit of $8\times10^{14}$~M$_\odot$~\cite{bbbo96}
\end{itemize}

\subsection{Science from the SZ Effect}
\begin{figure}
   \includegraphics[width=2.1in]{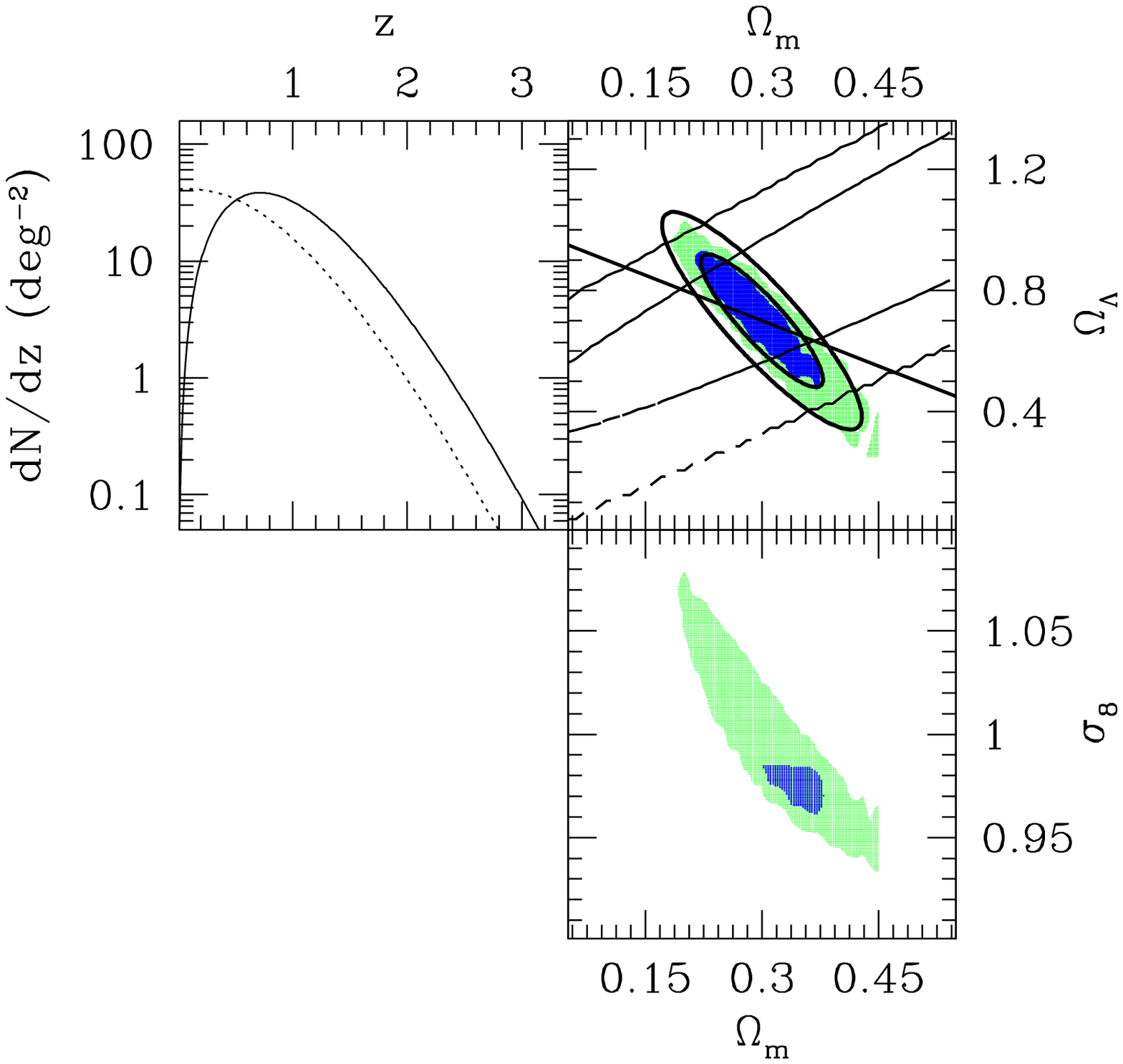}
   \includegraphics[width=2.1in]{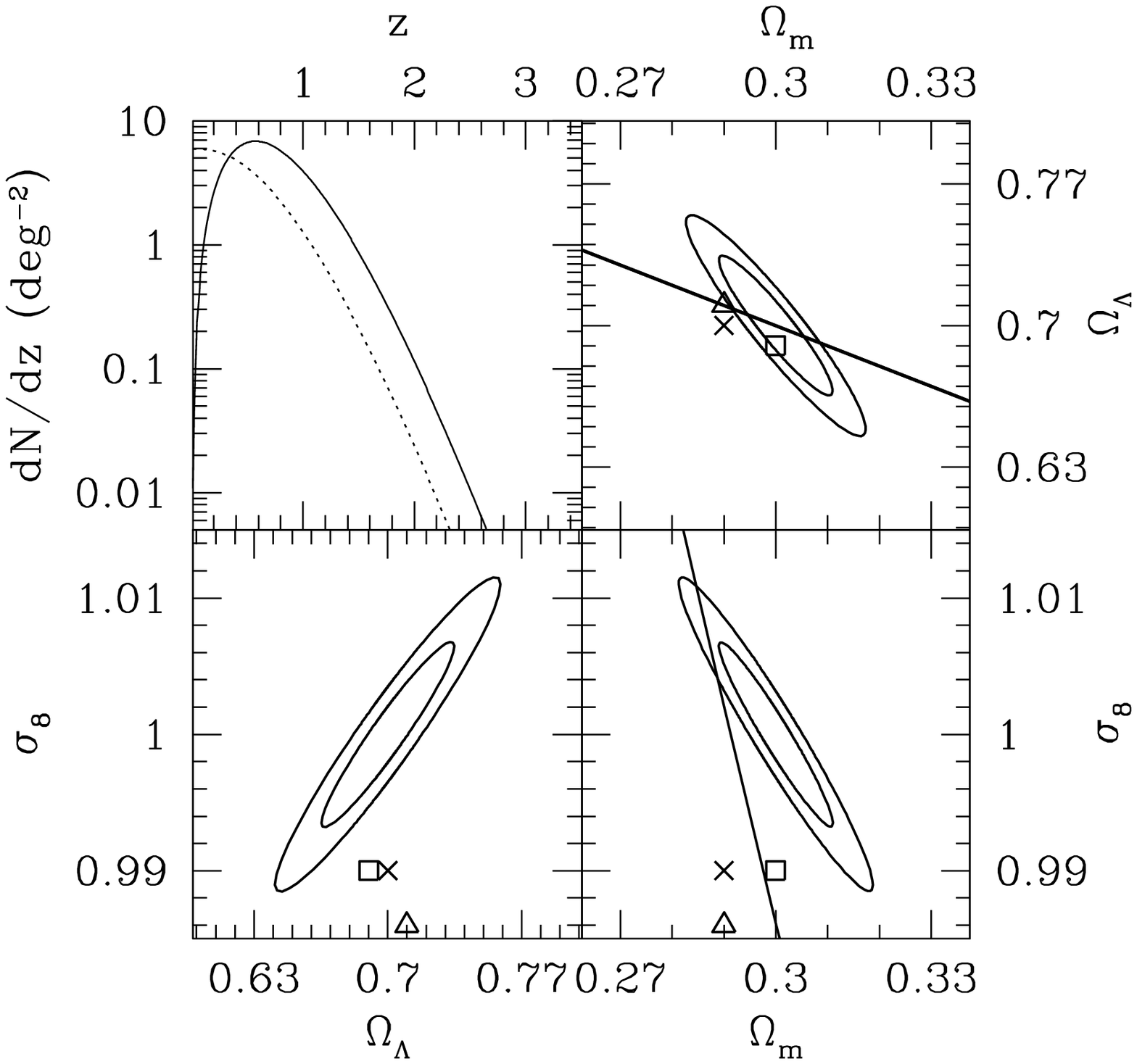}
   \includegraphics[width=2.1in]{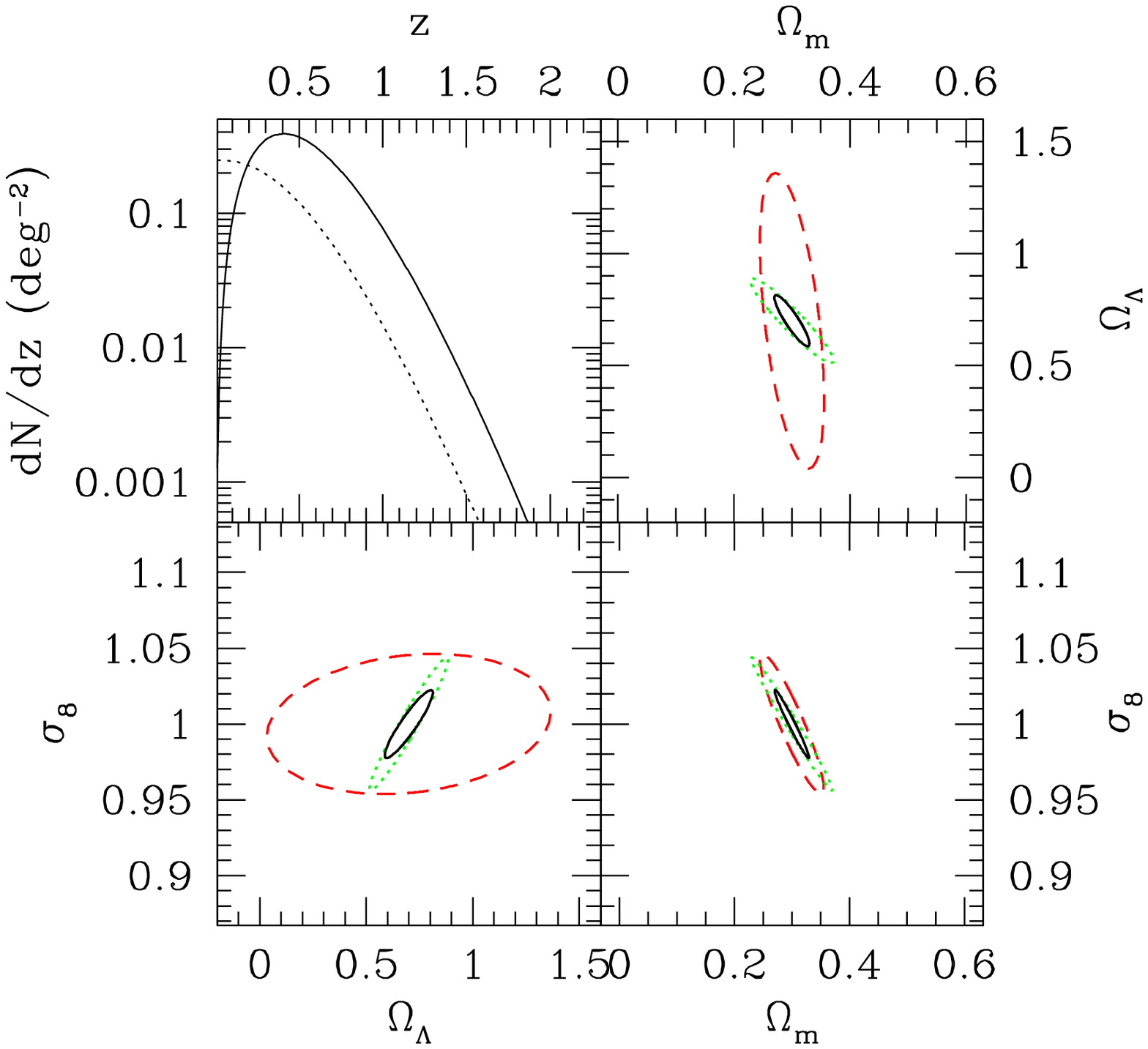}
   \caption[]{The ability of cluster surveys to probe cosmological parameters (figure from~\cite{hhm01},
courtesy of G. Holder). The upper left panel of each figure shows the
expected differential counts of clusters (the dashed line shows the
cumulative numbers).  The ellipses in the remaining panels show 68\% and
95\% confidence limits. The surveys are, from left to right, deep, medium
deep and shallow, as defined in Section~\ref{sec-sz-spros}.}
   \label{sec-sz-fig2}
\end{figure}

Fig.~\ref{sec-sz-fig2}, from~\cite{hhm01}, simulates the limits that such
SZ surveys will set on a variety of cosmological parameters (see
also~\cite{ws98, hmh01}).

\subsubsection{Using the SZ effect to Probe Dark Energy}
\label{sec-sz-s1}
\begin{figure}
   \includegraphics[width=2.7in]{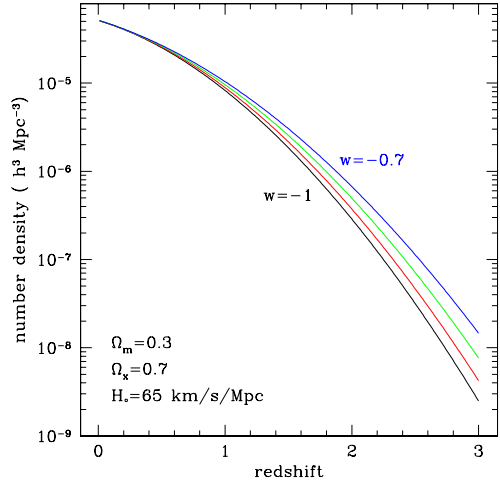}
   \caption[]{Number of clusters with $M>10^{14}$~M$_\odot$ as a function of
redshift for different values of $w$ (figure courtesy of G. Holder)}
   \label{sec-sz-fig3}
\end{figure}

Because the number density of clusters is sensitive to the epoch at which
cluster formation ends, this statistic constrains the allowable value of
$w$. Fig.~\ref{sec-sz-fig3} shows the dependence of the number density of
clusters with $M>10^{14}$~M$_\odot$ on $z$ for different values of $w$.
The variation is caused by the dependence of the size of a volume element
on $w$. The ability of cluster counts on the sky to constrain $w$ is
explored more fully in~\cite{hmh01} who show that the constraints are
orthogonal to those provided by SNIa measurements. The number density
depends on the integrated growth of structure, effectively integrating
expansion from time zero to the present day. In contrast, SN 1a
observations measure the angular diameter distance which is a function of
integral of expansion from present day back to the cluster.

The SZ effect can also be used in conjunction with X-ray measurements of a
cluster to probe the dependence of angular diameter distance, $d_A$, on
redshift.  The X-ray surface brightness of a cluster is given by:
\begin{equation}
S_x=\frac{1}{4\pi(1+z)^4}\int n_e n_H \Lambda_{eH} {\rm d}l
\end{equation}
where $n_H$ is the number density of hydrogen atoms in the ICM and
$\Lambda_{eH}$ is the X-ray cooling function integrated over the observing
band appropriate for the X-ray observation. By combining a measurement of
$y_{\rm th}$, which depends on the integral of $n_e$ along a line of sight
through the cluster, with measurements of the X-ray surface brightness,
which depends on the integral of $n_e^2$ along the same line of sight, the
linear depth of the cluster gas can be deduced. If the cluster is assumed
to be spherically symmetric, then by comparing the measured depth with the
angular extent of the cluster gas on the sky, the angular diameter
distance to the cluster and thus $H_0$ can be determined. The accuracy
from a single cluster is very low because of asphericity -- clusters are
not spherical as is assumed in the $d_A$ calculation. This can be
corrected for by observing many clusters at the same redshift.
Observations of 20-30 clusters, yields an estimate of
$H_0=63\pm3$~km\,s$^{-1}$\,Mpc$^{-1}$ (assuming statistical errors only)
~\cite{carl00}. A second source of uncertainty is clumping of the hot gas.
If the clumping factor $C=<n_e^2>^{1/2}/<n_e>$, then the Hubble constant
is overestimated by a factor $C^2$~\cite{carl00}.  Consequently, a better
understanding of the gas will be necessary to reliably determine angular
diameter distances.

Assuming that clumping issues can be understood, a sample of at least 60
clusters~\cite{kjs01} at a range of redshifts will determine 
$d_A(z)$ to a precision similar to that which has been obtained from SN1a
measurements. Such samples are within reach of existing experiments and
will be easily obtained by planned cluster surveys, with the caveat that
X-ray and optical follow-up measurements must also be obtained (see
Section~\ref{sec-sz-sout}).

\subsubsection{Other Cosmological Parameters from SZ Surveys}
%\label{sec-sz-s2}
{\bf Using the SZ Effect to Probe Baryonic Matter} Observations of the SZ
effect complement other tracers of clusters physics such as weak
gravitational lensing, and X-ray measurements. Like X-ray measurements, SZ
measurements are sensitive to the amount of hot gas in the cluster.
However, because the X-ray surface brightness is proportional to $\int
n_e^2{\rm d}l$, the SZ effect is better able to trace cooler, low-density
gas in the outskirts of a cluster, or between clusters. Consequently, the
SZ effect is a important tracer of the distribution of baryonic matter in
the universe.

{\bf Using the SZ Effect to Probe Dark Matter} Separation of the two
components of the SZ effect allows a direct measurement of $v_{\rm pec}$.
For an isothermal plasma, the peculiar velocity can be derived from the SZ
effect as follows:
\[ \frac{v_{\rm pec}}{c} = \frac{y_{\rm kin}}{y_{\rm th}} \frac{kT_{\rm
e}}{m_{\rm e}c^2}. \]

In principle, the gravitational potential that reflects the distribution
of all matter, including non-baryonic dark matter, can be reconstructed
from the peculiar velocity field, which comprises the residual motions of
matter over and above the Hubble flow. Both the evolution of bulk flows
with redshift and the amplitude distribution of peculiar velocities are
also important observables that can be used to test cosmological models
(see~\cite{cd01} for a review).  For distances greater than $\sim 150
h^{-1}$ Mpc, uncertainties in distance measurements become too large to
allow peculiar velocities to be determined accurately using traditional
techiques which rely on distance determinations.  The
SZ effect provides us with a way to measure the peculiar velocity of
clusters relative to the CMB with a precision that is redshift
independent.

As is clear from Figure~\ref{sec-sz-fig1} however, the thermal and kinetic
components can be separated only by measurements at mm wavelengths, close
to the null of the thermal effect. In addition, the relatively small
amplitude of the kinetic effect requires high sensitivity in order to
measure the effect with precision.  The ultimate limit to the peculiar
velocity determination of a single cluster comes from intrinsic CMB
anisotropies which have a spectrum that is indistinguishable from the
kinematic SZ effect.  The limits are expected to be about
$200\times(0.01/\tau)$\,km\,s$^{-1}$ for an experiment with a beam size of
1-2$'$~\cite{1996MNRAS.279..545H}. However, this limit is {\em
independent} of redshift, unlike empirical determinations of peculiar
velocity.  The Planck Surveyor, which will survey the whole sky at
22--850\,GHz, will provide a peculiar velocity survey that can used to set
limits of $50$--$200$\,km\,s$^{-1}$ to the bulk flows of 100$h^{-1}$ Mpc
volumes of space~\cite{1997A&A...325....9A, 2001A&A...374....1A}.
Follow-up measurements using ground-based instruments with higher angular
resolution will be able to set similar limits to smaller regions of space,
although the experiment is likely to be very time consuming.  Again,
redshift information from optical follow-up is needed to fully exploit
peculiar velocity measurements.

\subsection{Outstanding Issues}
\label{sec-sz-sout}
\subsubsection{Limits of modeling}
\label{sec-sz-smod}
The use of cluster number counts to constrain cosmological parameters
assumes that the physics of cluster formation is understood.  As discussed
above, the simplest model of cluster formation assumes that clusters are
virialized objects whose collapse is described by gravitational collapse
alone, and the predictions of results from SZ surveys discussed here are
based on that assumption. In practice other mechanisms such as feedback
from galaxy formation, or mergers can modify the results, especially for
low-mass clusters~\cite{vhs01, hc01}. Simulations are becoming quite
sophisticated but there is still substantial disagreement between
different results, as shown in Fig.~\ref{sec-sz-fig2} which shows a range
of predictions for the SZ power spectrum.  As the sophistication of
numerical modeling improves, with the added constraint of multi-frequency
investigations of known clusters, these uncertainties should reduce.

\begin{figure}
   \includegraphics[width=3in]{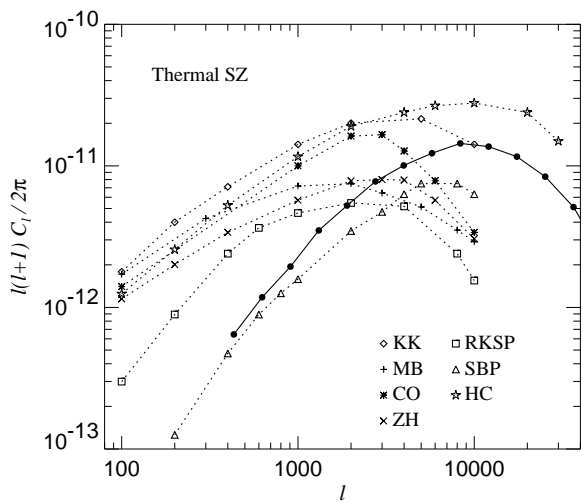}
   \caption[]{Simulations of the power spectrum of SZ thermal
fluctuations by several authors, from~\cite{2001ApJ...549..681S} (courtesy
of M. White).}
   \label{sec-sz-fig6}
\end{figure}

\subsubsection{Follow-up of Large Cluster Samples}
Future SZ surveys will provide a large cluster sample for which the SZ
flux, $y_{\rm th}$, and possibly $y_{\rm kin}$, will be determined. In
order to fully exploit the science from this sample, a well-planned
program of follow-up observations will be needed. The most critical
component will be optical follow-up to determine cluster redshifts.  This
will be a time-consuming task requiring large amounts of time on large
telescopes. In addition, X-ray measurements will be needed for programs to
determine $d_A(z)$ and peculiar velocities.

% figures should be put into the text as floats.
% Use the graphicx package (distributed with LaTeX2e).
% See the LaTeX Graphics Companion by Michel Goosens, Sebastian Rahtz,
% \begin{figure}
% %\includegraphics{}%
% \caption{}
% \label{}
% \end{figure}

% tables follow here or maybe be put in the text
% \begin{table}
% \caption{}
% \label{}
% \end{tabular}
% \end{table}

\begin{acknowledgments}
We would like the thank all the attendees who contributed to this
workshop, without whom we could not have written this report.  In
particular we would like to thank Peter Timbie, Sunil Golwala and
Gilbert Holder for generously allowing us to use materials from their
presentations, and for useful conversations after the conference.
Also we thank John Carlstrom, Wayne Hu and Martin White for
the use of their figures.  SEC acknowledges a Stanford Terman
Fellowship and support from NSF grants 9970797 and 9987360.  AHJ
acknowledges support from NASA LTSA Grant no.\ NAG5-6552 and NSF KDI
Grant no.\ 9872979 and from PPARC in the UK.  LK is supported by NASA LTSA
Grant no.\ NAG5-11098.

\end{acknowledgments}

%\bibliography{sec_refs,lek_refs,ahj_refs}
\bibliography{P4_3_CMBInflation}

\begin{thebibliography}{10}

\bibitem{linde00}
A. {Linde}, \physrep {\bf 333},  575  (2000).

\bibitem{guth00}
A.~H. {Guth}, \physrep {\bf 333},  555  (2000).

\bibitem{mukhanov92}
V. Mukhanov, H. Feldman, and R. Brandenberger, \physrep {\bf 215},  203
  (1992).

\bibitem{cormier99}
D. {Cormier} and R. {Holman}, \prd {\bf 60},  1301+  (1999).

\bibitem{dodelson97}
S. {Dodelson}, W.~H. {Kinney}, and E.~W. {Kolb}, \prd {\bf 56},  3207  (1997).

\bibitem{KolbTurner}
E.~W. {Kolb} and M.~S. {Turner}, {\em {The Early Universe}} (Frontiers in
  Physics, Addison-Wesley, Reading, MA, 1990).

\bibitem{Peebles}
P.~J.~E. {Peebles}, {\em {Principles of physical cosmology}} (Princeton Series
  in Physics, Princeton University Press, Princeton, NJ, 1993).

\bibitem{HuScoSil94}
W. {Hu}, D. {Scott}, and J. {Silk}, \prd {\bf 49},  648  (1994).

\bibitem{cmbfast}
U. {Seljak} and M. {Zaldarriaga}, \apj {\bf 469},  437  (1996).

\bibitem{camb}
A. {Lewis}, A. {Challinor}, and A. {Lasenby}, \apj {\bf 538},  473  (2000).

\bibitem{polprimer}
W. {Hu} and M. {White}, New Astronomy {\bf 2},  323  (1997).

\bibitem{maxiboom01}
A.~H. {Jaffe} {\it et~al.}, Physical Review Letters {\bf 86},  3475  (2001).

\bibitem{KamKosSte}
M. {Kamionkowski}, A. {Kosowsky}, and A. {Stebbins}, \prd {\bf 55},  7368
  (1997).

\bibitem{ZalSel97}
M. {Zaldarriaga} and U. {Seljak}, \prd {\bf 55},  1830  (1997).

\bibitem{cobe}
http://space.gsfc.nasa.gov/astro/cobe/.

\bibitem{Lee01}
A.~T. {Lee} {\it et~al.},   (2001), astro-ph/0104459.

\bibitem{netterfield01}
C.~B. {Netterfield} {\it et~al.},   (2001), astro-ph/0104460.

\bibitem{CBI}
S. {Padin} {\it et~al.}, \apjl {\bf 549},  L1  (2001).

\bibitem{halverson00}
N.~W. {Halverson} {\it et~al.},   (2001), astro-ph/0104489.

\bibitem{pryke01}
C. {Pryke} {\it et~al.},   (2001), astro-ph/0104490.

\bibitem{MAP}
http://map.gsfc.nasa.gov/.

\bibitem{Planck}
http://astro.estec.esa.nl/SA-general/Projects/Planck/.

\bibitem{StaggsGundrvw}
S. Staggs, J. Gundersen, and S. Chruch,  in {\em Microwave Foregrounds}, edited
  by A. de~Oliveira-Costa and M. Tegmark (ASP, ADDRESS, 1999).

\bibitem{POLAR01}
B.~G. {Keating} {\it et~al.}, \apjl {\bf 560},  L1  (2001).

\bibitem{PIQUE01}
M.~M. {Hedman} {\it et~al.}, \apjl {\bf 548},  L111  (2001).

\bibitem{DMRpol}
G. Smoot (unpublished).

\bibitem{coherence1}
J. {Magueijo}, A. {Albrecht}, P. {Ferreira}, and D. {Coulson}, \prd {\bf 54},
  3727  (1996).

\bibitem{ekpyrotic}
J. {Khoury}, B.~A. {Ovrut}, P.~J. {Steinhardt}, and N. {Turok},   (2001),
  hep-th/0103239.

\bibitem{wang01}
X. {Wang}, M. {Tegmark}, and M. {Zaldarriaga},   (2001), astro-ph/0105091.

\bibitem{kamionkowski00}
M. {Kamionkowski} and A. {Buchalter},   (2000), astro-ph/0001045.

\bibitem{knox00}
L. {Knox},  in {\em Particles, Strings and Cosmology} ({World Scientific},
  Singapore, 2000), p.\ 326.

\bibitem{knox01}
L. {Knox}, N. {Christensen}, and C. {Skordis},   (2001), astro-ph/0109232.

\bibitem{efstathiou01}
G. {Efstathiou} {\it et~al.},   (2001), astro-ph/0109152.

\bibitem{trotta01}
R. {Trotta}, A. {Riazuelo}, and R. {Durrer},   (2001), astro-ph/0104017.

\bibitem{croft00}
R. Croft {\it et~al.}, ApJ, submitted  (2000), astro-ph/0012324.

\bibitem{hannestad01}
S. Hannestad, S. Hansen, F. Villande, and A. Hamilton,   (2001),
  astro-ph/0103047.

\bibitem{zaldarriaga01}
M. {Zaldarriaga}, L. {Hui}, and M. {Tegmark}, \apj {\bf 557},  519  (2001).

\bibitem{forecast}
D.~J. {Eisenstein}, W. {Hu}, and M. {Tegmark}, \apj {\bf 518},  2  (1999).

\bibitem{wang99}
Y. {Wang}, D.~N. {Spergel}, and M.~A. {Strauss}, \apj {\bf 510},  20  (1999).

\bibitem{scgkt01}
S. Staggs {\it et~al.}, to appear in {\em report from Snowmass 2001: The Future
  of Particle Physics}  (2001).

\bibitem{hu01}
W. {Hu},   (2001), astro-ph/0108090.

\bibitem{verde01b}
L. {Verde} and D.~N. {Spergel},   (2001), astro-ph/0108179.

\bibitem{ZalSel98}
M. {Zaldarriaga} and U.~. {Seljak}, \prd {\bf 58},  3003  (1998).

\bibitem{LewChaTur01b}
A. {Lewis}, A. {Challinor}, and N. {Turok},   (2001), astro-ph/0108251.

\bibitem{BCJK99}
J.~R. Bond, R.~G. Crittenden, A.~H. Jaffe, and L.~E. Knox, Computers in Science
  and Engineering {\bf 1},  21  (1999).

\bibitem{BJK00}
J.~R. {Bond}, A.~H. {Jaffe}, and L. {Knox}, \apj {\bf 533},  19  (2000).

\bibitem{FerJafMNRAS00}
P.~G. {Ferreira} and A.~H. {Jaffe}, \mnras {\bf 312},  89  (2000).

\bibitem{Dore01}
O. {Dor{\' e}} {\it et~al.}, \aap {\bf 374},  358  (2001).

\bibitem{MADCAP}
http://www.nersc.gov/\~{}borrill/cmb/madcap.html.

\bibitem{stompor01Maps}
R. {Stompor} {\it et~al.},   (2001), astro-ph/0106451.

\bibitem{OSH99}
S.~P. {Oh}, D.~N. {Spergel}, and G. {Hinshaw}, \apj {\bf 510},  551  (1999).

\bibitem{DoreKnox01}
O. {Dore}, L. {Knox}, and A. {Peel}, \prd {\bf 64},  083001  (2001).

\bibitem{Hivon01}
E. {Hivon} {\it et~al.},   (2001), astro-ph/0105302.

\bibitem{ws98}
L. {Wang} and P.~J. {Steinhardt}, \apj {\bf 508},  483  (1998).

\bibitem{wco99}
L. Wang, R.~R. Caldwell, J.~P. Ostriker, and P.~J. Steinhardt, \apj {\bf 530},
  17  (2000).

\bibitem{2001ApJ...552....2G}
L. {Grego} {\it et~al.}, \apj {\bf 552},  2  (2001).

\bibitem{sz72}
R. Sunyaev and Y.~B. Zel'dovich, Comments Ap. Space Phys. {\bf 4},  173
  (1972).

\bibitem{cen99}
R. Cen and J. Ostriker, \apj {\bf 514},  1  (1999).

\bibitem{nsi00}
S. Nozawa, N. Itoh, Y. Kawana, and Y. Kohyama, \apj {\bf 536},  31  (2000).

\bibitem{reph95}
Y. Rephaeli, \apj {\bf 371},  L1  (1995).

\bibitem{1994ApJ...430L..13B}
N.~A. {Bahcall}, R. {Cen}, and M. {Gramann}, \apjl {\bf 430},  L13  (1994).

\bibitem{1996MNRAS.282..384M}
L. {Moscardini} {\it et~al.}, \mnras {\bf 282},  384  (1996).

\bibitem{bbbo96}
D. Barbosa, J.~G. Bartlett, A. Blanchard, and J. Oukbir, A\&A {\bf 314},  13
  (1996).

\bibitem{hc01}
G.~P. Holder and J.~E. Carlstrom, \apj {\bf 558},  515  (2001).

\bibitem{vhs01}
L. Verde, Z. Haiman, and D.~N. Spergel, \apj submitted  (2001),
  astroph/0106315.

\bibitem{carl00}
J.~E. Carlstrom {\it et~al.},  in {\em Constructing the Universe with Clusters
  of Galaxies}, edited by F. Durret and D. Gerbal (IAP 2000 meeting, Paris,
  France, 2000).

\bibitem{1990cmwb.book...77B}
M. {Birkinshaw},  in {\em ASSL Vol. 164: The Cosmic Microwave Backround: 25
  Years Later} (Kluwer Academic Publishers, Dordrecht, Netherlands, 1990), pp.\
  77--94.

\bibitem{rmc00}
E.~D. Reese {\it et~al.}, \apj {\bf 533},  38  (2000).

\bibitem{bc01}
B. Benson {\it et~al.} (unpublished).

\bibitem{cb01}
S. Church {\it et~al.} (unpublished).

\bibitem{daw01}
K.~S. {Dawson} {\it et~al.}, \apjl {\bf 553},  L1  (2001).

\bibitem{sub00}
R. {Subrahmanyan} {\it et~al.}, \mnras {\bf 315},  808  (2000).

\bibitem{sel96}
U. Seljak and M. Zaldarriaga, \apj {\bf 469},  437  (1996).

\bibitem{1999mifo.conf..199H}
G.~P. {Holder} and J.~E. {Carlstrom},  in {\em ASP Conf. Ser. 181: Microwave
  Foregrounds} (ASP, San Francisco, 1999), pp.\ 199+.

\bibitem{2001ApJ...549..681S}
V. {Springel}, M. {White}, and L. {Hernquist}, \apj {\bf 549},  681  (2001).

\bibitem{carl98}
J. Carlstrom {\it et~al.}, Physica Scripta, volume T {\bf 85},  148  (2000).

\bibitem{kjs01}
R. Kneissl {\it et~al.}, \mnras  (2001), astro-ph/0103042.

\bibitem{hhm01}
G. Holder, Z. Haiman, and J.~J. Mohr, \apj {\bf 560},  L111  (2001).

\bibitem{hmh01}
Z. Haiman, J.~J. Mohr, and G.~P. Holder, Ap. J. {\bf 553},  545  (2001).

\bibitem{cd01}
S. Courteau and A. Dekel,  in {\em Astrophysical Ages and Time Scales} (ASP
  Conference Series, San Francisco, 2001), p.\ in press.

\bibitem{1996MNRAS.279..545H}
M.~G. {Haehnelt} and M. {Tegmark}, \mnras {\bf 279},  545  (1996).

\bibitem{1997A&A...325....9A}
N. {Aghanim} {\it et~al.}, \aap {\bf 325},  9  (1997).

\bibitem{2001A&A...374....1A}
N. {Aghanim}, K.~M. {G{\' o}rski}, and J.-L. {Puget}, \aap {\bf 374},  1
  (2001).

\end{thebibliography}

\end{document}